\documentclass[journal=jacsat,manuscript=article]{achemso}
\usepackage[utf8]{inputenc}

\usepackage[dvipsnames]{xcolor}
\usepackage[utf8]{inputenc}

\usepackage[T1]{fontenc}
\DeclareUnicodeCharacter{2212}{\ensuremath{-}}   %
\DeclareUnicodeCharacter{2013}{\textendash}      %
\DeclareUnicodeCharacter{2014}{\textemdash}      %

\usepackage{newtxtext,newtxmath}

\usepackage{graphicx}
\usepackage{amsmath}

\usepackage{booktabs}

\usepackage{url}
\usepackage[hidelinks]{hyperref}

\usepackage{siunitx}

\title{
Reconstruction-Dependent Imaging, Reactivity and Local Reduction of the CeO$_2$(100) surface
}

\author{Kyungmin Kim}
\altaffiliation{These authors contributed equally.}
\affiliation[Osaka University]
{Osaka University, 1-3 Machikaneyama, Toyonaka, 560-8531, Japan}

\author{Manuel Gonz\'{a}lez Lastre}
\altaffiliation{These authors contributed equally.}
\affiliation[UAM]
{Departamento de F\'{i}sica Te\'{o}rica de la Materia Condensada, Universidad Aut\'{o}noma de Madrid, 28049 Madrid, Spain}

\author{Estefan\'{i}a Fern\'{a}ndez-Villanueva}
\affiliation[UAM]
{Departamento de F\'{i}sica Te\'{o}rica de la Materia Condensada, Universidad Aut\'{o}noma de Madrid, 28049 Madrid, Spain}
\alsoaffiliation[CSIC]
{Instituto de Cat\'{a}lisis y Petroleoqu\'{i}mica (ICP-CSIC), Calle de Marie Curie 2, 28049 Madrid, Spain}
\alsoaffiliation[UPV]
{Universitat Polit\`{e}cnica de Val\`{e}ncia, Cam\'{i} de Vera s/n, 46022 Val\`{e}ncia, Spain} 

\author{Pablo Pou}
\affiliation[UAM]
{Departamento de F\'{i}sica Te\'{o}rica de la Materia Condensada, Universidad Aut\'{o}noma de Madrid, 28049 Madrid, Spain}
\alsoaffiliation[IFIMAC]
{Condensed Matter Physics Center (IFIMAC), Universidad Aut\'{o}noma de Madrid, 28049 Madrid, Spain}

\author{Hossein Sepehri-Amin}
\affiliation[NIMS]
{National Institute for Materials Science (NIMS), 1-2-1 Sengen, Tsukuba, 305-0047, Japan}

\author{Masayuki Abe}
\affiliation[Osaka University]
{Osaka University, 1-3 Machikaneyama, Toyonaka, 560-8531, Japan}

\author{Shigeki Kawai}
\affiliation[NIMS]
{National Institute for Materials Science (NIMS), 1-2-1 Sengen, Tsukuba, 305-0047, Japan}

\author{M. Ver\'{o}nica Ganduglia-Pirovano}
\affiliation[CSIC]
{Instituto de Cat\'{a}lisis y Petroleoqu\'{i}mica (ICP-CSIC), Calle de Marie Curie 2, 28049 Madrid, Spain}

\author{Ruben Perez}
\email{ruben.perez@uam.es}
\affiliation[UAM]
{Departamento de F\'{i}sica Te\'{o}rica de la Materia Condensada, Universidad Aut\'{o}noma de Madrid, 28049 Madrid, Spain}
\alsoaffiliation[IFIMAC]
{Condensed Matter Physics Center (IFIMAC), Universidad Aut\'{o}noma de Madrid, 28049 Madrid, Spain}

\author{Oscar Custance}
\email{custance.oscar@nims.go.jp}
\affiliation[NIMS]
{National Institute for Materials Science (NIMS), 1-2-1 Sengen, Tsukuba, 305-0047, Japan}

\usepackage{xr-hyper}

\makeatletter
\newcommand*{\addFileDependency}[1]{%
  \typeout{(#1)}%
  \@addtofilelist{#1}%
  \IfFileExists{#1}{}{\typeout{No file #1.}}%
}
\makeatother
\newcommand*{\myexternaldocument}[1]{%
  \externaldocument{#1}%
  \addFileDependency{#1.tex}%
  \addFileDependency{#1.aux}%
}
\myexternaldocument{SI}

\begin{document}

\maketitle

\begin{abstract}
The possibility of mapping the local reactivity and reduction state to the atomic structure of chemically active oxide surfaces opens new avenues for further understanding of catalysis.
Here, we combine scanning tunnelling (STM) and atomic force microscopy (AFM) with first-principles modelling to explore this possibility on the CeO2(100) surface.
While STM reveals the periodicity of cerium-terminated and oxygen-terminated CeO$_2$(100) reconstructions coexisting on the same surface, AFM imaging and force spectroscopy provide direct identification of the exposed atomic species and their reactivity as the chemical interaction with the probe.
Density functional theory based STM and AFM simulations reproduce the main experimental observations and show that STM contrast cannot be in general assigned to the atomic positions of certain chemical species, as traditionally assumed from previous studies.
Simulated STM contrast of the two reconstructions across different reduction states associated with the removal of oxygen atoms in deeper layers, evidence that STM alone does not offer a robust fingerprint of the local reduction state for the cerium-terminated reconstruction, but it is sensitive to the reduced state in the case of the oxygen-terminated one, being able to provide information on a mixed distribution of Ce$^{3+}$ and Ce$^{4+}$ ions on the first sub-surface Ce layer.
\end{abstract}

\section{Introduction}

Cerium dioxide (CeO$_{2}$, ceria) is a pivotal material in heterogeneous catalysis, used to accelerate reactions that require oxygen exchange,
which is facilitated by the unique redox properties of this oxide \textemdash involving the creation of surface active sites by the Ce$^{4+}$~$\rightleftharpoons$~Ce$^{3+}$ switching\textemdash\space and its 
remarkable oxygen storage and release capacity~\cite{Tuller1979Feb, Skorodumova2004Feb, Castleton2007Dec}.
These functionalities make CeO$_{2}$ essential in important catalytic applications, such as hydrogen production via the water-gas-shift reaction, methanol conversion, and carbon monoxide oxidation~\cite{5,6,7}.
It is therefore of utmost interest to investigate the CeO$_{2}$ surfaces where these reactions occur and characterize their structure, defects, reactivity and reduction state
down to the atomic scale.~\cite{doi:10.1021/jacsau.5c00095}.

The CeO$_{2}$(111) surface has been the primary focus of fundamental reactivity studies due to its stability and accessibility as thin-films
~\cite{8,9,10,11}.
In contrast, the CeO${_2}$(100) facet provides a more challenging but highly relevant model system: its intrinsic polarity~\cite{15}, charge compensation, and variety of reconstructions~\cite{16,17,18,19, Zhang_STME_CeO2_100_4x6} are associated with distinct physical and chemical properties of interest for catalytic chemistry~\cite{Hu2022Apr,Capdevila-Cortada2017Mar,Mullins2012Sep}; yet the availability of experimental studies on this surface is still scarce~\cite{Skorodumova2004Feb,13,14}.

Scanning tunneling microscopy (STM) has traditionally been the method of choice for the characterization of the atomic-scale structure and defects of CeO$_{2}$ surfaces~\cite{Esch2005,NiliusVeronica,LustembergNiliusAuDiffusionOnGold}.
STM imaging of nanocubes exposing the (100) facet revealed a (2$\times$2) surface periodicity with small patches of a c(2$\times$2) structure~\cite{Niklas}, and a CeO$_4$-pyramidal (2$\times$2) model was proposed as the most stable structure based on density functional theory (DFT) calculations~\cite{21}.
As-grown diluted CeO$_{2}$(100) and CeO$_{2}$(111) mixed samples showing both the (2$\times$2) and the c(2$\times$2) reconstructions for the (100) facet were also investigated using STM~\cite{STETSOVYCH2013766}.
These studies have established the characteristic surface periodicity of these reconstructions and motivated atomistic models, but the exposed atomic species at the surface, the precise atomic-site assignment of the STM signal, and the local reduction state of the reconstructed domains have remained difficult to assess~\cite{Niklas,STETSOVYCH2013766}.
The applicability of STM to the study of the CeO$_2$(100) surfaces is limited, not only by the intrinsic wide band-gap character of this oxide, but also by the strong structural corrugation and the bias-dependent contribution of Ce- and O-derived states of this surface.
Atomic force microscopy (AFM)~\cite{Giessibl95-Science} and site-specific force spectroscopy~\cite{Lantz01-Science-ForceSpectroscopy,Sugimoto07-Nature-ChemicalIdentification} operated at the atomic scale can overcome the limitations of STM providing an identification of the surface chemical species in bare metal-oxide surfaces~\cite{Stetsovych2015Jun,DieboldSetvin_ForceSpectroscopy} and  in the presence of relevant adsorbates~\cite{oscar_manuel_ceria_111_water, KimCe3plus}. However, this chemical insight is essentially limited to the top surface layer, leaving out crucial information regarding the structure and local reduction state in complex, highly-corrugated surface reconstructions.

In this work, we demonstrate that a combination of STM and AFM imaging with site-specific force spectroscopy using copper-oxide-functionalized probes~\cite{Monig2018Aug,OxygenTerminationOfACuOxTip} together with DFT-based simulations for these different information channels
provides details on the atomic-scale structure, the reactivity of the exposed chemical species and the local reduction state on complex oxide surfaces.
We illustrate the performance of this combined approach with the characterization of the  (2$\times$2) and  c(2$\times$2) reconstructions  found at CeO$_2$(100) islands grown on a Cu(111) substrate, putting on a firm basis the pyramid-terminated CeO$_4$(100)  and the oxygen-terminated O-t(100) models proposed in  previous CeO$_2$(100) studies, where the corresponding atomic structures were inferred mainly from STM symmetry and apparent corrugation~\cite{Niklas,STETSOVYCH2013766}.
While STM identifies the characteristic periodicity reported for these reconstructions~\cite{Niklas,STETSOVYCH2013766}, AFM imaging and force spectroscopy enable direct assignment of the atomic species exposed at the surface: both Ce and O sites are resolved on CeO$_4$-t(100), whereas the top-layer oxygen atoms are directly imaged for the O-t(100) domains.
These experiments, together with first-principles calculations to simulate (i) site-specific 
force spectroscopy curves, (ii) constant-height AFM images, the latter using the Full Density Based Model (FDBM)~\cite{Ellner2019, Ventura-Macias2023Oct}, and (iii)  STM topography images show that the STM contrast (e.g. STM maxima) cannot be in general assigned to the atomic positions of certain chemical species, as traditionally assumed from previous studies on the non-polar CeO$_2$(111) surface.  
Furthermore, we analyze the corresponding simulated STM contrast of the CeO$_4$-t(100) and O-t(100) surfaces across different reduction states associated  with the removal of O atoms in deeper layers that are not accessible to AFM imaging.
Our simulations evidence that  STM alone does not offer a robust fingerprint of the local reduction state for the CeO$_4$-t(100), as different oxidation states and Ce$^{3+}$ distributions can yield similar apparent contrast in this highly corrugated  surface geometry. However, STM is able to reveal the reduced state of the O-t(100), and the presence of a mixed distribution of Ce$^{3+}$ and Ce$^{4+}$ sites on the first underlying Ce layer.

Our combined experimental-theoretical approach establishes a framework that connects the surface periodicity observed by STM with the atomic-site assignment obtained by AFM, the local response measured by AFM spectroscopy, and the information on the reduction state gathered from STM, providing a powerful tool for the atomic-scale characterization of the structure and local chemical reactivity for highly corrugated oxide reconstructions.

\section*{Experimental and computational details}\label{Methods}

\subsubsection*{Scanning probe microscopy experiments}\label{Exp}

An ultra-high vacuum (UHV) system equipped with a home-made scanning probe microscope connected to a bath cryostat was used to carry out the experiments.
STM and AFM measurements were performed at 4.8~K using the KolibriSensor (SPECS, Germany) and the Nanonis SPM Control System (SPECS, Germany). 
The probe was initially sharpened ex situ to a typical apex radius of $\sim$15~nm using a focused ion beam and conditioned for atomic resolution imaging in areas of the copper oxide surface that coexists with the CeO$_{2}$(100) islands.
Frequency modulation detection with constant oscillation amplitude~\cite{Albrecht91-J.Appl.Phys.-FreqModulation} was deployed to operate the force sensor, using the shift in the free-oscillation resonant frequency ($\Delta f$) as the primary signal for AFM detection.
Spectroscopic measurements were carried out by approaching the sample towards the probe (fixed in our experimental setup) a specific distance from the imaging conditions, and recording the $\Delta f$ signal upon retracting the sample until reaching the free-oscillation regime of the AFM sensor (a $\Delta f(z)$ curve).
During STM acquisition, the bias voltage was applied to the sample.
The CeO$_{2}$(100) islands were grown on a Cu(111) surface following a procedure similar to that described in O. Stetsovych et al.~\cite{STETSOVYCH2013766}.

\subsubsection*{Density Functional Theory Calculations}

Structure optimizations and simulated force-distance calculations were performed using Density Functional Theory (DFT) as implemented in the VASP code (version 5.4.4)~\cite{Kresse1993Jan,Kresse1994May,Kresse1996Oct,Kresse1996Jul,Kresse1999Jan} with the slab-supercell approach~\cite{Payne1992Oct}. 
The projector augmented wave (PAW) method~\cite{Blochl1994Dec} was used to describe the valence electrons of the atomic species: Ce (4f, 5s, 5p, 5d, 6s), O (2s, 2p) and H (1s), with a plane-wave cutoff energy of 415~eV. 
The DFT+U approach proposed by Dudarev et al.~\cite{Dudarev1998Jan} was applied to properly localize the electrons on the Ce$^{3+}$ ions of the support, using a $U_{eff}$ value of $U - J = 4.5$~eV for the Ce 4f electrons~\cite{Fabris}.
We also implemented the generalized gradient approximation (GGA) PBE exchange-correlation functional by Perdew, Burke, and Ernzerhof (PBE)~\cite{Perdew1996Oct}. 
The van der Waals (vdW) dispersion energy correction was considered through the DFT-D3 approach~\cite{Grimme2010Apr,Grimme2011May}.

\subsection{Structure calculations}

\noindent 
The CeO$_4$-t(100) and O-t(100) surface reconstructions were modeled using p(2$\times$2) and c(2$\times$2) structures, respectively, and
consist of ten and nine atomic layers. %
The centered %
c(2$\times$2) cell is non-primitive and rotated 45$^\circ$ with respect to the primitive p(2$\times$2) structure. Therefore, the ($\sqrt{2}\times\sqrt{2}$)R45$^\circ$ primitive of the c(2$\times$2) cell is equivalent to that of the p(2$\times$2) rotated 45$^\circ$ (Fig.~S3D)

For the fully oxidized stoichiometric surfaces, the c(2$\times$2) O-t(100) structure is the O-terminated CeO$_2$(100) configuration known as the checkerboard model. In this structure,
only half of the oxygen atoms (eight in our c(2$\times$2) cell) occupy the topmost layer, while the remaining half (another eight atoms) are positioned at the bottom of the slab to compensate for polarization (Fig.~\ref{Fig1}F).
The CeO$_4$-t(100) p(2$\times$2) structure was built by manually adding one CeO$_2$ unit to a p(2$\times$2) cell of the checkerboard model. The unit is added between two surface oxygen atoms perpendicularly to the line that connects them, thus generating the CeO$_4$ pyramid-like feature characteristic of this reconstruction (Fig. \ref{Fig1}E). Note that this mixed termination is the 50\% Ce-terminated CeO$_2$(100) surface denoted as (100)-Ce$_{50\%}$ in our recent publication~\cite{Arturo2025}.

For the sub-stoichiometric reduced surfaces, we used the B2 and C2 models by Pan et al.~\cite{Niklas} for O-t(100) and CeO$_4$-t(100), respectively. To that end, starting from the oxidized c(2$\times$2) O-t(100) structure, the reduced c(2$\times$2) O-t(100) surface is built by removing one of every two consecutive oxygen atoms at the top layer, four in total, leaving the other four spread out %
preserving the symmetry (Fig.~S3D).
Similarly, to obtain the reduced p(2$\times$2) CeO$_4$-t(100) surface, the two topmost surface oxygen atoms that do not belong to any pyramid in the oxidized p(2$\times$2) CeO$_4$-t(100) structure were removed (Fig.~S3B).
Since each oxygen atom accommodates two electrons, two Ce$^{3+}$ ions are produced per oxygen atom removed. In the reduced p(2$\times$2) CeO$_4$-t(100) surface, the pyramidal Ce atoms become Ce$^{3+}$ ions, together with three out of the four second layer Ce atoms. For the c(2$\times$2) O-t(100) reduced surface, two relative positions for the eight Ce$^{3+}$ ions produced are considered (Fig.~S3D).
Finally, an oxidized p(2$\times$4) CeO$_4$-t(100) model was also built to study adsorbates (Fig. \ref{Fig7}).

In all cases, the first bottom layers for O and Ce atoms were kept fixed at their optimized bulk-truncated positions during geometry optimization, whereas the rest of the atoms were allowed to fully relax. According to the Monkhorst--Pack method~\cite{Monkhorst1976Jun}, a (2$\times$2$\times$1), (3$\times$3$\times$1) and (3$\times$1$\times$1) k-point mesh was used to sample the Brillouin zone for the c(2$\times$2), p(2$\times$2) and p(2$\times$4) surfaces, respectively.
For the gas-phase calculations of the molecules, a (15$\times$15$\times$15) $\mathring{\text{A}}$$^3$ cell was employed, with $\Gamma$-point only.

Adsorption energies on the CeO$_4$-t(100) surface were calculated as:
\
\begin{equation}
\Delta E_{ads} = E(\text{Molecule}/\text{CeO$_4$-t(100)}) - E(\text{Molecule}) - E(\text{CeO$_4$-t(100)})
\end{equation}
\
where $E(\text{Molecule}/\text{CeO$_4$-t(100)})$ is the energy of the structure with the molecule adsorbed on the surface, $E(\text{Molecule})$ is the energy of the molecule in the gas phase, and $E(\text{CeO$_4$-t(100))}$ is the total energy of the clean model surface.

\noindent

\subsection{Atomic Force Microscopy simulations}

\subsubsection*{Probe model}

We modeled the experimental Cu$_x$O-terminated probe using a rigid CO molecule with its oxygen atom oriented toward the surface. 
This simple model reproduces the short-range Pauli repulsion between the lone pair of the oxygen-terminated apex and the sample, while avoiding an explicit atomistic description of the full Cu$_x$O probe structure~\cite{oscar_manuel_ceria_111_water}.
We supplemented this model with a long-range background term describing the van der Waals contribution from the mesoscopic part of the probe, obtained by fitting the experimental $\Delta f(z)$ curves as described below.
This approach %
provided excellent results for the simulations of $\Delta f(z)$ curves on the CeO$_2$(111) surface, yielding quantitative agreement with the experiment.~\cite{oscar_manuel_ceria_111_water}.  The strengths and limitations of this probe model and our failed efforts to find a more realistic apex structure are presented in section S3.2 of the Supplementary Information.

\subsubsection{Simulated frequency shift curves}

$\Delta f(z)$ curves were computed from DFT-obtained force curves calculated by vertically approaching a rigid CO molecule towards the corresponding surface site, summing up the vertical force components on both C and O atoms of the molecule at each step height.
The forces were converted to $\Delta f$ using classical perturbation theory~\cite{GiessiblPRB1997} as implemented in the formalism proposed by Giessibl~\cite{Giessibl2001APL}. For this conversion, we used the experimental values of $f_0$ and $A$ (Figs.~\ref{Fig4} and~\ref{Fig5}) and the stiffness $k$ from the manufacturer specifications of the KolibriSensor~\cite{KolibriDevelopmentPaper}.
The long-range vdW background originating from the mesoscopic part of the probe was determined by fitting the tails of the experimental $\Delta f(z)$ curves to the analytical expression to describe the vdW interaction of a sphere over a plane~\cite{Lantz01-Science-ForceSpectroscopy, Garcia2002Sep}.
\begin{equation} 
F{_\mathrm{vdW}}(z) = \frac{C}{z^{2}}
\Rightarrow
\Delta f_{\mathrm{vdW}}(z) = \frac{f_0}{\sqrt{8 \pi} \cdot k \cdot A^{3 / 2}} \cdot \frac{C}{\left(z\right)^{3 / 2}}, 
\label{eq:giessibl}
\end{equation} 
Fitting the experimental $\Delta f(z)$ curves over the long--range regime on CeO$_4$-t(100) and O-t(100) surfaces (Figs.~\ref{Fig4} and~\ref{Fig5}) provides a single best--fit value of $C=-1.4311\times10^{-6}$~N\,pm$^{2}$. The background $\Delta f_{\mathrm{vdW}}(z)$ from Eq.~\eqref{eq:giessibl} was then added to every DFT-obtained $\Delta f(z)$ curve and to the AFM image simulations described below.

\subsubsection{Simulated constant-height AFM images}

Constant-height AFM images were generated with the FDBM~\cite{Ellner2019}, which evaluates the total probe-surface interaction energy ($V_{\mathrm{total}}$) for an inert probe \textemdash such as a CO molecule\textemdash\space as the sum of short-range (SR) Pauli repulsion, electrostatics (ES) and the vdW dispersion correction as modeled by the same semiempirical D3-vdW method~\cite{Grimme2010Apr} deployed in our DFT-force calculations: $V_{\mathrm{total}} = V_{\mathrm{SR}} + V_{\mathrm{ES}} + V_{\mathrm{D3-vdW}}$.
We calculated the $V_{\mathrm{SR}}$ and $V_{\mathrm{ES}}$ energies using the probe and the surface charge densities ($\rho_{\mathrm{probe}}$, $\rho_{\mathrm{surface}}$) and the surface electrostatic potential $\Phi_{\mathrm{surface}}$ obtained from independent DFT calculations:
\[
V_{\mathrm{SR}} \;=\; V_0 \int \!\big[\rho_{\mathrm{probe}}(\mathbf{r})\,\rho_{\mathrm{surface}}(\mathbf{r})\big]^{\alpha}\, d\mathbf{r}, 
\qquad
V_{\mathrm{ES}} \;=\; \int \!\rho_{\mathrm{probe}}(\mathbf{r})\,\Phi_{\mathrm{surface}}(\mathbf{r})\, d\mathbf{r}.
\]
We applied the same short-range parameters validated for the partially reduced CeO$_{2}$(111) surface in our previous work~\cite{oscar_manuel_ceria_111_water}, ($\alpha=1.08$, $V_0=36.96~\mathrm{eV/}$ $\mathring{\text{A}}$ $^{3(2\alpha-1)}$).
At each lateral position in a constant-height scan, the total force was obtained by differentiating $V_{\mathrm{total}}$, and then adding the long-range probe-surface vdW force from Eq.~\eqref{eq:giessibl}. 
The resulting force field was converted to $\Delta f$ using the same perturbative framework~\cite{GiessiblPRB1997, Giessibl2001APL} and the experimental $f_0$, $A$ and $k$ parameters employed for simulating the $\Delta f(z)$ curves.
The implementation followed the open-source \texttt{DBSPM} repository (\href{https://github.com/SPMTH/DBSPM}{\texttt{github.com/SPMTH/DBSPM}}).

\subsection{STM Simulations}

As the experimental Cu$_x$O probe is oxygen-terminated, we assume that tunneling is dominated by a probe-apex state of $\sigma$ symmetry. Under this assumption, STM images were simulated with the Tersoff--Hamann approximation~\cite{TersoffHamann}, where the tunneling current $I$ at a given bias voltage $V$ is proportional to the integrated LDOS of the sample at the probe position, 
\[
I(\mathbf{r}, V) \propto \int_{E_F}^{E_F+eV} \mathcal{D}(\mathbf{r}, E)\, dE ,
\]
with $E_F$ the Fermi energy and $\mathcal{D}(\mathbf{r}, E)$ the LDOS of the sample at position $\mathbf{r}$ and energy $E$ . 

The integration windows were selected from the calculated electronic structure to probe the characteristic contributions of the two atomic species, taking the projected density of states (pDOS) for each system as a reference. The corresponding pDOS are displayed in Figs.~S14~to~S17.
Empty-state simulations were obtained by integrating from $E-E_F=0$ to $+1.5$~eV for all surface models, encompassing the predominantly Ce-derived unoccupied states. For filled-state simulations, the integration window was chosen within the occupied O-derived manifold. In the reduced surfaces, the upper limit was adjusted for each model to avoid the localized occupied Ce$^{3+}$ states present close to the Fermi level. %
The resulting energy windows are summarized in Table~\ref{tab:stm_windows}.

The integrated LDOS was obtained from the partial charge densities calculated with VASP. Constant-height images correspond to spatial maps of the integrated LDOS evaluated at a fixed probe-sample separation, whereas constant-current images were generated from isosurfaces of the integrated LDOS. The same energy windows were used for the STM images and for the three-dimensional LDOS isosurfaces employed to analyze the origin of the STM contrast in the Supporting Information.

\begin{table}[h!]
\centering
\caption{
\textbf{Energy integration windows used for the STM simulations and LDOS isosurfaces.}
Energies are given relative to the Fermi level, $E_F$. Empty-state windows were selected to probe predominantly Ce-derived unoccupied states, whereas filled-state windows were chosen within the O-derived occupied manifold while excluding the localized Ce$^{3+}$ states close to $E_F$ in the reduced models. The labels for the O-t(100) models denote distributing %
the Ce$^{3+}$ ions between the first and second Ce layers (mix), and locating all of them at the top Ce layer (top), respectively (see Figure~S3D).}
\label{tab:stm_windows}
\begin{tabular}{lcc}
\toprule

Surface model & Empty states (eV) & Filled states (eV) \\
\midrule
Oxidized CeO$_4$-t(100) & $0$ to $+1.5$ & $-3.0$ to $-1.0$ \\
Reduced CeO$_4$-t(100)  & $0$ to $+1.5$ & $-3.0$ to $-1.75$ \\
Reduced O-t(100)-mix    & $0$ to $+1.5$ & $-3.0$ to $-2.0$ \\
Reduced O-t(100)-top    & $0$ to $+1.5$ & $-3.0$ to $-2.0$ \\

\bottomrule
\end{tabular}
\end{table}

\section{Results and discussion}

\subsection{The CeO$_4$-t(100) reconstruction}

\begin{figure}[b!]
\centering
\includegraphics[width=1.0\textwidth]{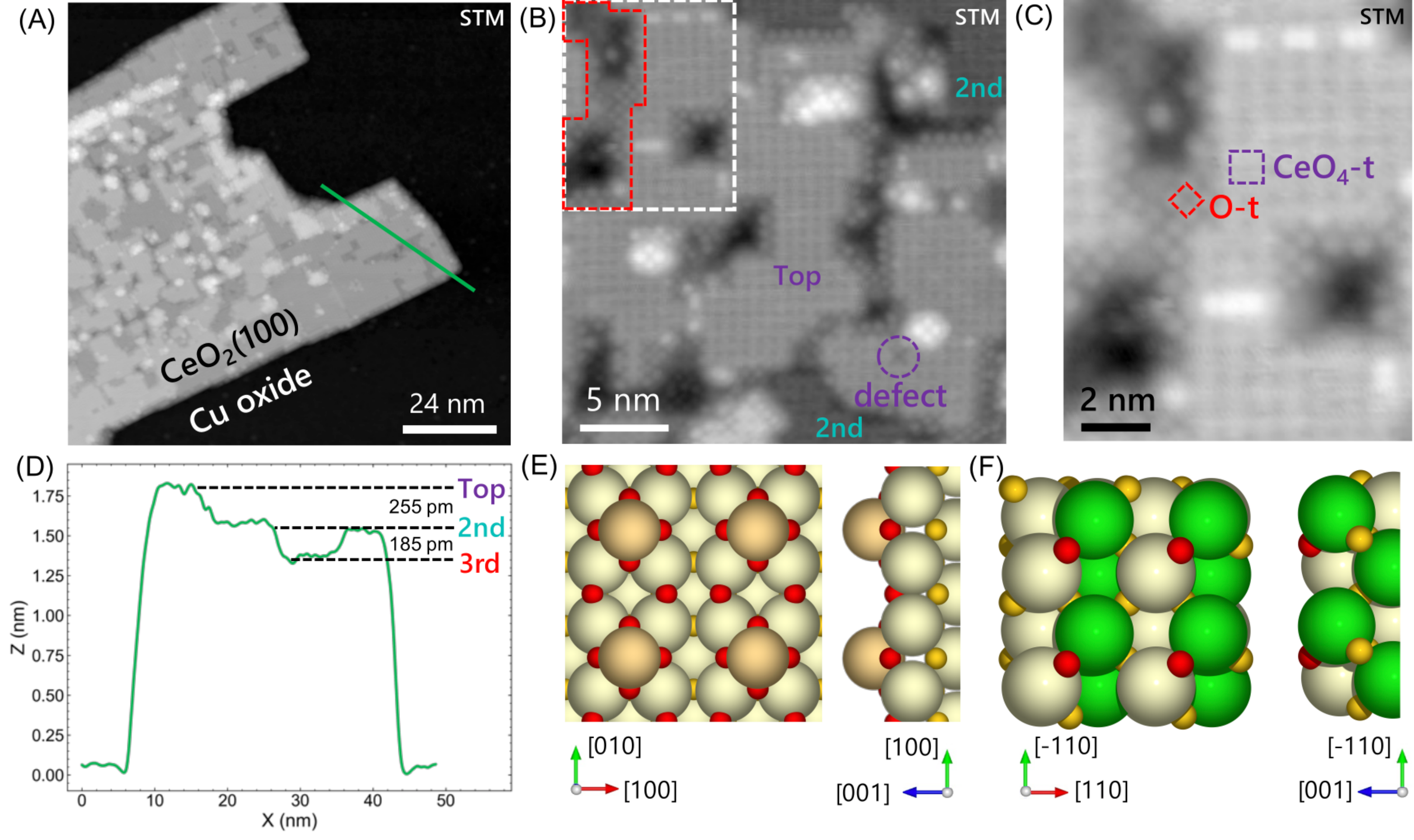} 
\caption{%
\textbf{Ceria (100) morphology and surface reconstructions.}
(\textbf{A}) Topographic STM image of a CeO$_2$(100) island grown on pre-oxidized Cu(111).
(\textbf{B}) Topographic STM image of the top and second terraces of the island showing the predominance of the (2$\times$2) reconstruction, CeO$_4$-t(100) for short.
A region displaying the c(2$\times$2) reconstruction, O-t(100) for short, is enclosed by dashed red lines.    
(\textbf{C}) Zoom of the region marked by a white square in (B), featuring CeO$_4$-t(100) and O-t(100) domains with different unit-cell orientations.
(\textbf{D}) Height profile along the trace in (A) displaying representative step heights between terraces.
(\textbf{E} and \textbf{F}) DFT-optimized structures of the CeO$_4$-t(100) and O-t(100) terminations, respectively.
These models serve as reference structures used throughout the manuscript. 
Alternative reduced configurations and Ce$^{3+}$ ion distributions are discussed in Figures~S3~and~S4.
Colors: surface oxygen (red), subsurface oxygen (yellow), Ce$^{3+}$ ions (green), surface Ce$^{4+}$ ions (sand), and pyramid Ce$^{4+}$ ions (beige).
Acquisition parameters: $V=2.7$~V and $I=2$~pA.
}
\label{Fig1} 
\end{figure}

Figure~\ref{Fig1}A depicts a topographic STM image of a $\sim$1.7~nm thick CeO$_{2}$(100) island.
The surface displays the characteristic tetragonal formations of the CeO$_{2}$(100) facet.
A profile in Fig.~\ref{Fig1}D shows an apparent atomic step height of 255~pm between terraces, consistent with the expected 270~pm theoretical value for the CeO$_2$(100) surface~\cite{STETSOVYCH2013766} considering the variation of the effective tunnelling gap with the CeO$_2$ thickness.
The island is predominantly exposing the top and second terraces \textemdash indicative of layer-by-layer growth \textemdash\space with occasional patches of a third terrace 185~pm below the second one (Fig.~\ref{Fig1}D).
Figure~\ref{Fig1}B presents a topographic STM image of an area of the island that displays patches of the top and second terraces.
Both surfaces exhibit a well-ordered square atomic arrangement (Fig.~\ref{Fig1}B) with lattice parameters of 750~pm (top) and 780~pm (second), close to the 740~pm expected for the CeO$_4$-t(100) reconstruction of the CeO$_2$(100) surface~\cite{Niklas} (Fig.~\ref{Fig1}E).
Areas decorating the step edges between the top and second terraces show, however, small patches of the O-t(100) reconstruction, with a unit %
cell rotated $45^{\circ}$ relative to the CeO$_4$-t(100) surface (Figs.~\ref{Fig1}C~and~F) and a lattice constant of 560~pm, consistent with previous reports~\cite{Niklas,STETSOVYCH2013766}.
The O-t(100) reconstruction is also found at the third terrace highlighted in the profile in Fig.~\ref{Fig1}D. Bright protrusions on the top terrace mark the onset of an additional layer, while domain boundaries, pits exposing lower terraces, and isolated atomic-scale defects are also visible (Figs.~\ref{Fig1}B~and~\ref{Fig1}C).
The presence of these point defects provides useful information for analysing the atomic structure and chemical nature of the surface.

\begin{figure}
\centering
\includegraphics[width=1.0\textwidth]{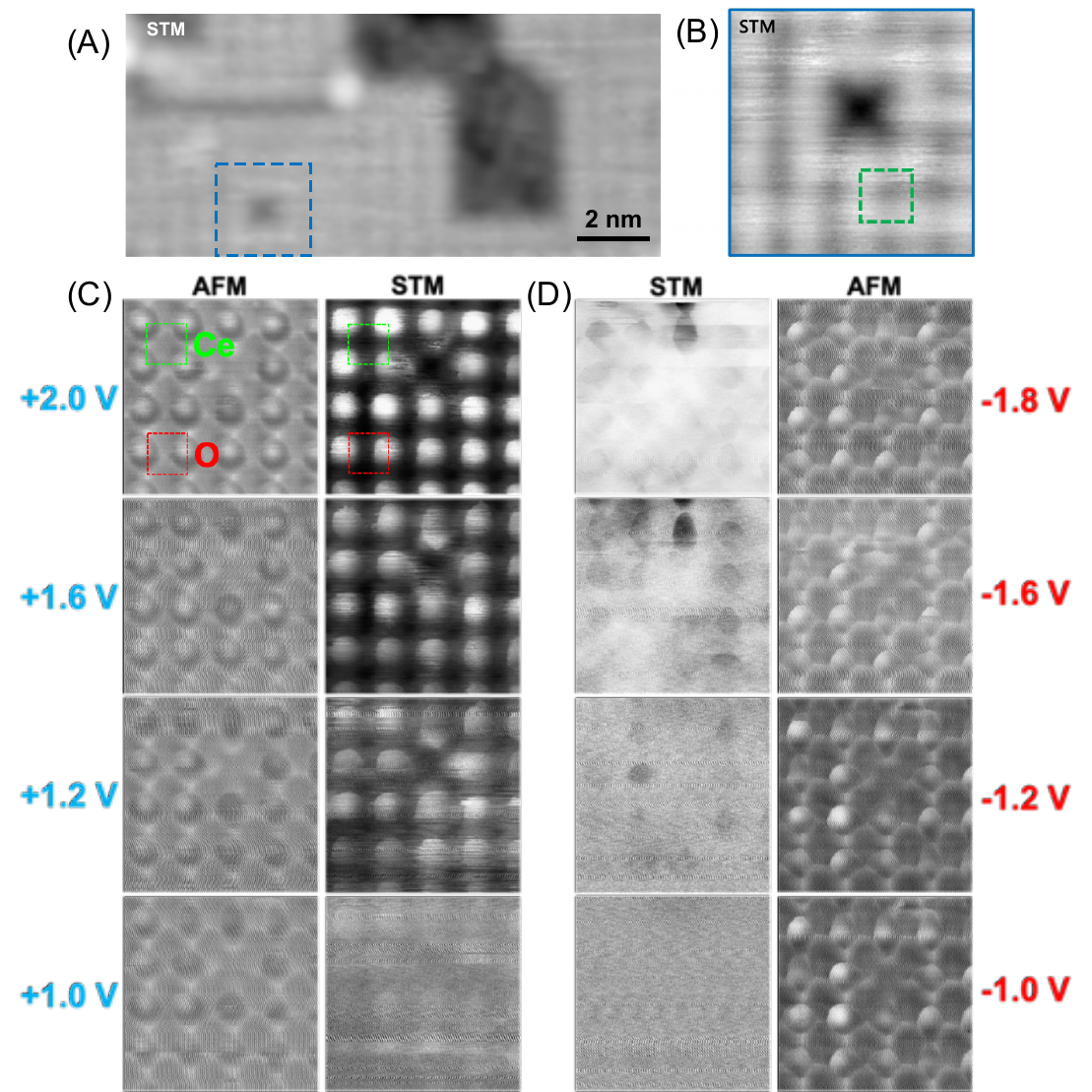} 
\caption{%
\textbf{Atomic assignment by bias-dependent, simultaneous constant-height STM and AFM imaging of the CeO$_4$-t(100) reconstruction.}
(\textbf{A}) Topographic STM image of a CeO$_4$-t(100) region of the island exhibiting native point defects.
(\textbf{B}) Detail of the defect marked in (A).
(\textbf{C}) and (\textbf{D}) Simultaneously acquired constant-height bias-dependent AFM and STM images to investigate the empty (C) and filled (D) states of the surface.
In empty-state STM, the bright protrusions are related to the Ce sites, whereas AFM signal resolves the outermost Ce and O atomic positions across all bias voltages explored.
Acquisition parameters: $V=2.7$~V, $I=2$~pA, for (A) and (B); $A=60$~pm, and $f_0=992175$~Hz for AFM imaging.
Data at $-1.6$ and $-1.8$~V were recorded retracting 20~pm from the surface to prevent probe modifications.
}
\label{Fig2} 
\end{figure}

Figure~\ref{Fig2}A shows an area of the island where the CeO$_4$-t(100) is predominant.   
This area exhibits point defects, which appear in the STM topography as faint cross-like features (Fig.~\ref{Fig2}B), as well as linear arrays of them~\cite{Niklas}
(Fig.~S1).
Topographic STM alone does not resolve the atomic structure of the CeO$_4$-t(100) surface; it only provides a general view of its square lattice symmetry~\cite{Niklas, Stetsovych2015Jun}. 
Simultaneous constant-height STM and AFM imaging at different sample bias values enables, however, visualization and chemical assignment of the atomic species (Figs.~\ref{Fig2}C~and~D). 
This improvement in resolution arises from the higher proximity of the probe towards the surface required for atomic-resolution AFM imaging, and the absence of the limitations imposed by the topographic STM feedback in wide band gap oxides.

\begin{figure}[b!]
\centering
\includegraphics[width=1.0\linewidth]{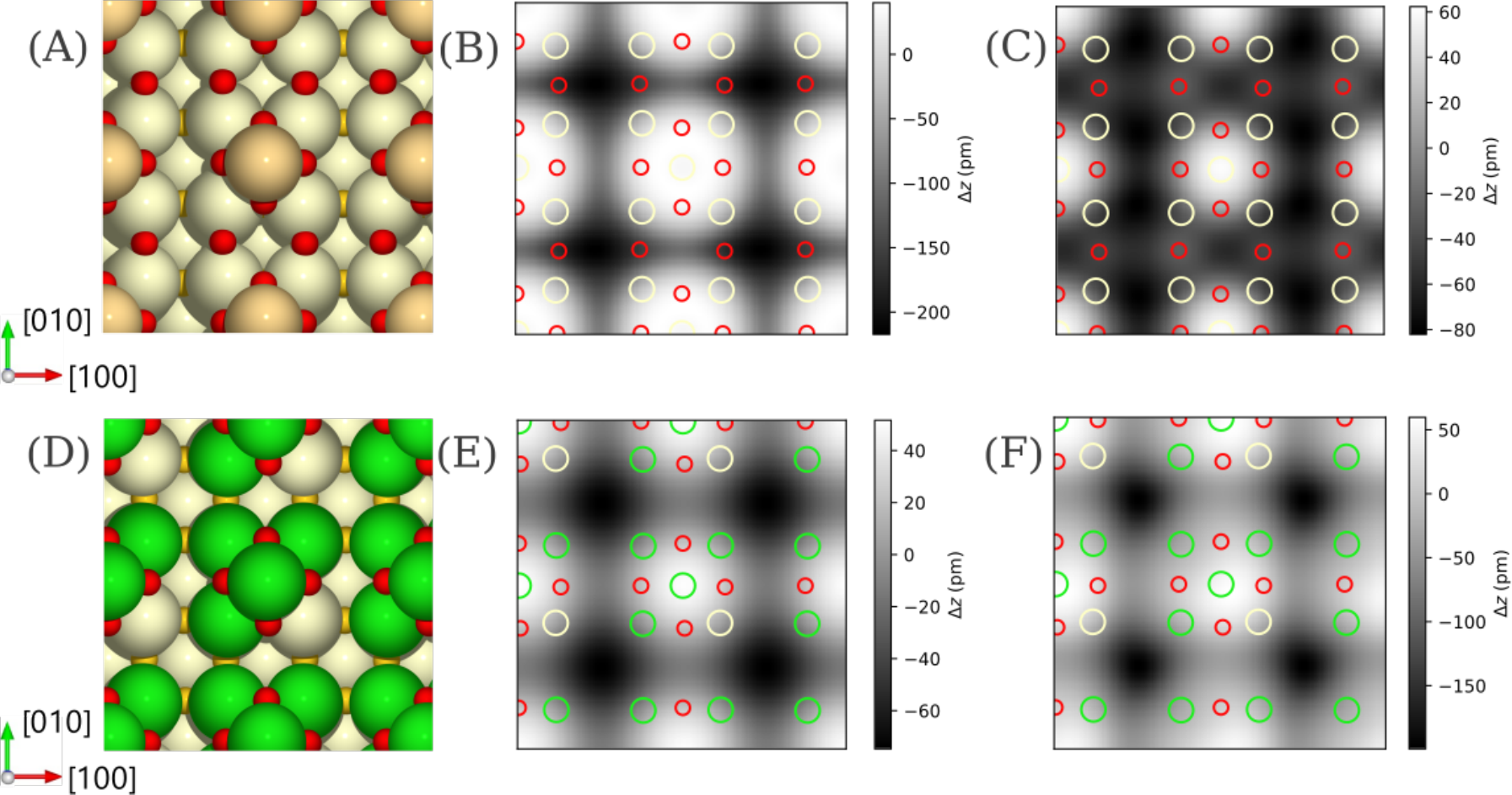}
\caption{%
\textbf{Simulated constant-current STM images of oxidized and reduced CeO$_4$-t(100).}
(\textbf{A}, \textbf{D}) Structural models of the oxidized and reduced terminations, respectively.
(\textbf{B}, \textbf{C}) Empty- and filled-state STM images of the oxidized surface.
(\textbf{E}, \textbf{F}) Empty- and filled-state STM images of the reduced surface.
The CeO$_4$ pyramid is centered in the simulation cell in all STM calculations.
Isovalues correspond to the mean isovalue in a plane 3~$\mathring{\mathrm{A}}$ above the top Ce atom on the CeO$_4$ pyramid. The empty- and filled-state isovalues are $1.06\times10^{-4}$ and $1.62\times10^{-5}$~states~$\mathring{\mathrm{A}}^{-3}$ for the oxidized surface, and $9.42\times10^{-5}$ and $2.01\times10^{-5}$~states~$\mathring{\mathrm{A}}^{-3}$ for the reduced surface, respectively.
Removing the oxygen atoms between pyramids \textemdash leading to the reduced model\textemdash\space just produces subtle changes in the contrast of the simulated STM images.
}
\label{Fig3}
\end{figure}

Under the conventional interpretation of ceria empty-state STM images ($V>0$), the bright protrusions in the STM images of Fig.~\ref{Fig2}C would be assigned to Ce$^{4+}$ sites~\cite{NiliusVeronica,Castleton2007Dec,Wolf2019Jun_ceria_stm_simulations,KimCe3plus} and could therefore be taken as evidence of an oxidized termination.
The simulations presented below show, however, that this interpretation is not reliable for the strongly corrugated CeO$_4$-t(100) surface.
The simultaneously acquired AFM images resolve the Ce positions ascribed to a bright contrast in the STM signal and, in addition, reveal signatures of the oxygen atoms.
At the defect site, the absence of the Ce-related STM protrusion together with the AFM response \textemdash four bright spots slightly elongated toward the center of the defect\textemdash\space call for a Ce vacancy and underlying oxygen atoms.

In filled-states STM images ($V<0$), atomic resolution degrades and only a faint contrast over some atomic sites is detected (Fig.~\ref{Fig2}D).
On the contrary, the AFM retains contrast on both species, with the oxygen atoms appearing as small bright protrusions between the signal ascribed to the Ce atoms.
These measurements confirm the CeO$_{2}$(100)-(2$\times$2) reconstruction exhibiting a CeO$_4$ pyramidal termination.

\subsubsection{Origin of the STM contrast in the CeO$_4$-t(100) surface}

The STM contrast of ceria surfaces is governed by the electronic structure of Ce and O states~\cite{NiliusVeronica,Castleton2007Dec,Wolf2019Jun_ceria_stm_simulations,KimCe3plus, Esch2005, Wolf2019Jun_ceria_stm_simulations}
In empty-state imaging, the tunneling current is mainly associated with unoccupied Ce$^{4+}$ states, %
while Ce$^{3+}$ states lie below the Fermi level and, due to their strong localization, typically do not contribute to the tunneling signal (Figure~S15)
For the low-corrugation %
CeO$_2$(111) surface, this different contribution to the STM contrast %
has been used to identify reduced regions and oxygen vacancies, as sites associated with Ce$^{3+}$ ions often appear as a suppressed or missing signal in empty-state STM images~\cite{Esch2005,Wolf2019Jun_ceria_stm_simulations}.
Conversely, filled-state STM contrast is primarily associated with occupied O-derived states.
This established idea suggests that STM could, in principle, distinguish between oxidized and reduced areas of the CeO$_4$-t(100) surface.
In the oxidized case, empty-state imaging should highlight the Ce$^{4+}$ ions of the CeO$_4$ pyramids, whereas in a reduced termination of the surface it is expected that the Ce$^{3+}$ sites would only weakly contribute to the STM signal. 
Filled-state imaging, in turn, should probe the O-derived occupied states. 

To assess whether this interpretation remains valid for the strongly corrugated CeO$_4$-t(100) surface, we simulated constant-current STM images for oxidized and reduced models of this reconstruction within the Tersoff--Hamann approximation (Fig.~\ref{Fig3}).
For the oxidized model, the empty-state simulation follows the expected behavior, with the dominant contrast centered on the Ce atoms of the CeO$_4$ pyramids. 
Surprisingly, the reduced model yields a very similar contrast for the empty-state case. 
Although the states of the Ce$^{3+}$ ions at the pyramids have negligible projected weight in the empty-state energy window, the integrated LDOS develops a pronounced maximum in the vacuum region above the geometric position of the Ce$^{3+}$ sites (Figure~S15E).
At the probe-sample distances relevant for STM imaging, the tunneling current is therefore dominated by this maximum in the vacuum region 
that has no significant contribution from the atom-projected Ce$^{3+}$ density of states.
In addition, isosurface values are not too different between the oxidized and reduced surfaces.
As a result, the oxidized and reduced CeO$_4$-t(100) terminations are practically indistinguishable in the simulated empty-state STM images.
This shows that the absence of Ce$^{4+}$ ions in the top layer calls for a reassessment of the conventional interpretation of STM images.

The filled-state simulations, for both oxidized and reduced models, confirm that the dominant occupied contribution originates from O-derived states (Figures~S14~and~S15). 
However, the apparent contrast remains centered around the CeO$_4$ pyramids instead of directly resolving the oxygen sub-lattice. 
This behavior reflects the spatial overlap and vacuum decay of the occupied-state density, together with the large vertical corrugation of the CeO$_4$-t(100) reconstruction. 
Moreover, the filled-state tunneling current is much weaker than in empty-state imaging, consistent with the poor atomic resolution observed experimentally in the filled-state STM images of Fig.~\ref{Fig2}D. 
The absence of atomic-resolution filled-states STM images in previous works on this surface supports these findings.

The STM simulations reveal only subtle differences between the oxidized and reduced CeO$_4$-t(100) surfaces, featuring small variations in the shape and intensity of the contrast above the CeO$_4$ pyramids and a weak contribution from the additional oxygen atoms present in the oxidized case.
These differences are unlikely to provide a robust experimental fingerprint to identify oxidized and reduced areas of the surface, since they are expected to be further attenuated by the finite spatial extent and electronic structure of the experimental probe. 
Thus, although the projected electronic structure of Ce$^{4+}$, Ce$^{3+}$, and O-derived states follows the trends known from studies on the CeO$_2$(111) surface (Figs.~S14~and~S15),
the larger structural corrugation of CeO$_4$-t(100) changes how these states contribute to the tunneling current. 
Therefore, the STM contrast of this reconstruction cannot be interpreted solely in terms of atomic energy levels; it also requires consideration of the vacuum decay of the integrated LDOS and of the height differences between the atomic sublattices. 
We conclude that STM alone is not sufficient to unambiguously determine the local oxidation state of the CeO$_4$-t(100) domains.

\subsubsection{AFM contrast and reactivity of the CeO$_4$-t(100) surface}

\begin{figure}
\centering
\includegraphics[width=0.9\textwidth]{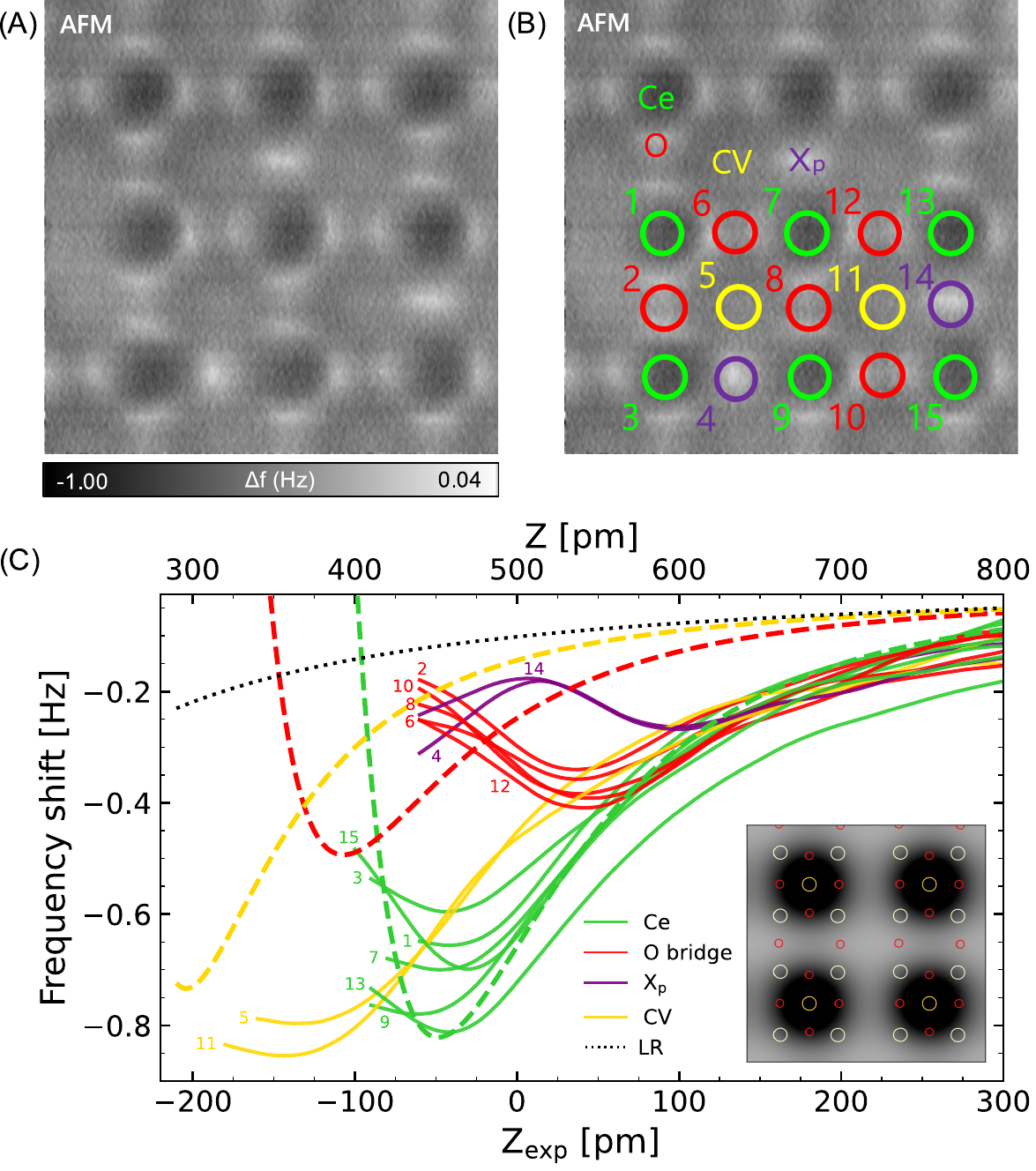} 
\caption{%
\textbf{Site-specific force spectroscopy on the CeO$_4$-t(100) surface.}
(\textbf{A}) Constant-height AFM image of a CeO$_4$-t(100) region.
(\textbf{B}) Same image with labels for the spectroscopy acquisition sites. Color code: green for cerium atoms (Ce); red for bridge positions mid-distance between two oxygen atoms of adjacent CeO$_4$ pyramids; purple for those displaying a single bright protrusion (X$_p$) at similar surface location as the bridge site; and yellow for a coordination vacancy (CV) site.
(\textbf{C}) Experimental $\Delta f(z)$ curves (solid) and DFT-based simulations (dashed) for the sites marked in (B). The black dotted line represents the long range (LR) van der Waals contribution of the mesoscopic part of the probe (see Materials and Methods).
Inset: Constant-height $\Delta f$ map of a CeO$_4$-t(100) region calculated using FDBM setting a probe-surface separation of Z=~450~pm. The image main feature is a deep attractive interaction over the top Ce extending over the CeO$_4$ pyramid (see Figure S6).
Acquisition parameters: $V=0$~V, $A=60$~pm, and $f_0=992175$~Hz.
}
\label{Fig4} 
\end{figure}

The lack of resolution over the O sites in occupied-states STM images discussed in the previous section is overcome by the AFM, which can resolve both Ce and O atoms at the topmost positions of the CeO$_4$-t(100) surface. 
In an attempt to gain further insight into the atomic features resolved by the AFM, we deployed force spectroscopic measurements~\cite{Lantz01-Science-ForceSpectroscopy}, in which the probe-surface interaction is recorded over several surface sites upon changing the probe-sample separation from a common origin, resulting in a set of $\Delta f(z)$ curves. 
When exploring different atomic positions with an identical probe termination, variations among the corresponding $\Delta f(z)$ curves contain chemical information about the surface atomic species~\cite{Sugimoto07-Nature-ChemicalIdentification, DieboldSetvin_ForceSpectroscopy}. In particular, they provide information on the local reactivity of the surface atoms toward the probe apex.

Figure~\ref{Fig4}A shows an atomically resolved constant-height AFM image of the CeO$_4$-t(100) surface acquired with a different probe apex and probe-surface separation than those used in Fig.~\ref{Fig2}.
The image displays depressions at the Ce sites on top of the CeO$_4$ pyramids and bright spots around them,
corresponding to the positions where the oxygen atoms of the CeO$_4$ pyramids are expected to be located (Fig.~\ref{Fig1}E).
The features observed in the AFM image point towards an oxygen-terminated probe, consistent with our conditioning procedure~\cite{OxygenTerminationOfACuOxTip, oscar_manuel_ceria_111_water} (see Materials and Methods).
Force spectroscopy curves were recorded over representative surface sites, marked in the AFM image displayed in Fig.~\ref{Fig4}B.
At the imaging height used to acquire Fig.~\ref{Fig4}A, the Ce atoms appear as depressions, reflecting a strong attractive interaction with the probe.
This statement is confirmed by the $\Delta f(Z_{\mathrm{exp}})$ curves measured over the Ce sites (green in Figs.~\ref{Fig4}B~and~C), which display deep attractive minima close to the imaging height (origin in the distance axis).
Although the minima registered over the Ce atoms occur at nearly the same probe-surface separation, small variations in the depth of the minima and in the slope of the curves toward the repulsive interaction region indicate modest changes in the local atomic environment.
The coordination vacancy between the CeO$_4$ pyramids (yellow in Figs.~\ref{Fig4}B~and~C) exhibits a minimum of comparable magnitude to the top Ce atoms, but shifted by $\sim$100~pm toward smaller $Z_{\mathrm{exp}}$ values. This shift is consistent with the local geometrical depression of the vacancy and with the contribution of nearby subsurface atoms to the  interaction with the probe.
The red curves in Fig.~\ref{Fig4}C were measured at bridge positions between adjacent top Ce atoms and capture the interaction of the probe with two oxygen atoms of adjacent CeO$_4$ pyramids that produce the bright spots observed in the image.
Despite these oxygen atoms lying $\sim$100~pm below the top Ce atom (Figure~S3
), the $\Delta f (Z_{\mathrm{exp}})$ curves measured over these bridge sites reach their minima $\sim$100~pm above the position of the Ce minima, indicating an earlier onset of a repulsive interaction.

In Fig.~\ref{Fig4}A, two types of contrast are observed at the bridge regions between neighboring top Ce atoms: one displaying a two-lobed feature (red circles in Fig.~\ref{Fig4}B), which we assign to two oxygen atoms belonging to adjacent CeO$_4$ pyramids; and another class of contrast, labeled X$_{p}$ (purple in Fig.~\ref{Fig4}B), which instead appears as a single bright oval.
 These X$_p$ sites exhibit a distinct $\Delta f(Z_{\mathrm{exp}})$ curve, suggesting that they correspond to a local perturbation at the pristine CeO$_4$-t(100) surface. Their atomistic origin is analyzed in a separate section below.

To rationalize this set of force spectroscopy measurements, we compared the experimental $\Delta f(Z_{\mathrm{exp}})$ curves with DFT-based simulations using the rigid CO probe model described in Methods. 
This model provides a simplified representation of the oxygen-terminated Cu$_\mathrm{x}$O apex and has previously yielded quantitative agreement for the less corrugated CeO$_2$(111) surface~\cite{oscar_manuel_ceria_111_water}.
For the top Ce atoms of the CeO$_4$ pyramids, the calculated curve reproduces reasonably well both the shape and the depth of the experimental minima (Fig.~\ref{Fig4}C). 
Aligning the calculated and experimental minima at these Ce sites gives the correspondence $Z_{\mathrm{exp}}=0 \Leftrightarrow Z\approx500$~pm, where the DFT reference $Z=0$ corresponds to the plane defined by the top Ce atoms.
The long-range vdW contribution of the mesoscopic part of the probe is included in this comparison as described in Methods.
At comparable imaging heights, the constant-height $\Delta f$ map calculated using FDBM reproduces the dominant attractive contrast over the top Ce atoms (inset of Fig.~\ref{Fig4}C).

For atomic sites located deeper in the surface, the agreement in the comparison of simulated and experimental curves is less satisfactory; particularly, the calculated minima for the oxygen-bridge and coordination vacancy positions occur at closer distances towards the surface than in the experiment (Fig.\ref{Fig4}C).
This discrepancy points towards a limitation of the relatively simple rigid-CO model to describe a Cu$_\mathrm{x}$O probe apex.
For the CeO$_4$-t(100) surface, the FDBM decomposition of interactions (Fig.~S6)
shows that the AFM contrast at the relevant distance of the experiment is largely governed by electrostatic interactions rather than by purely short-range Pauli repulsion.
Because electrostatic forces are more sensitive to the extended charge distribution of the probe apex, a model based on the description of the forefront atom of the probe alone does not capture in full the experimental interaction at the lower-lying oxygen-bridge and coordination vacancy sites.
The top Ce atoms are less affected by this limitation because their higher position 
localizes the interaction closer to the forefront atom of the probe.
A detailed analysis of the force decomposition and alternative Cu$_\mathrm{x}$O probe models is provided in sections S3.1 and S3.2 of the Supplementary Information.

An open question is whether the AFM data can provide information about the local reduction state of the CeO$_4$-t(100) surface. 
In the oxidized model (Figs.~\ref{Fig1}E~and~\ref{Fig3}A), the surface contains rows of oxygen atoms between the CeO$_4$ pyramids that are missing in the reduced model (Fig.~\ref{Fig3}D).
These oxygen atoms lie $\sim$130~pm below the top Ce atoms and $\sim$30~pm below the oxygen atoms of the CeO$_4$ pyramids (Fig.~S3A).
The absence of a clear AFM signal over the sites associated with these rows of atoms could indicate local reduction of the surface. 
However, the lack of signal could also arise from the lower height of these oxygen atoms, which may prevent resolving them with a similar contrast as the oxygen atoms at the CeO$_4$ pyramids. Trying to experimentally get signal from these rows of oxygen atoms would compel us to scan at closer probe surface distances that might compromise the integrity of probe or surface.
Thus, while some AFM features are compatible with reduced regions, the present STM and AFM data do not allow a definitive assignment of the  oxidation-state of the CeO$_4$-t(100) surface.

\subsection{O-terminated (100) reconstruction}

\begin{figure}
\centering
\includegraphics[width=0.95\textwidth]{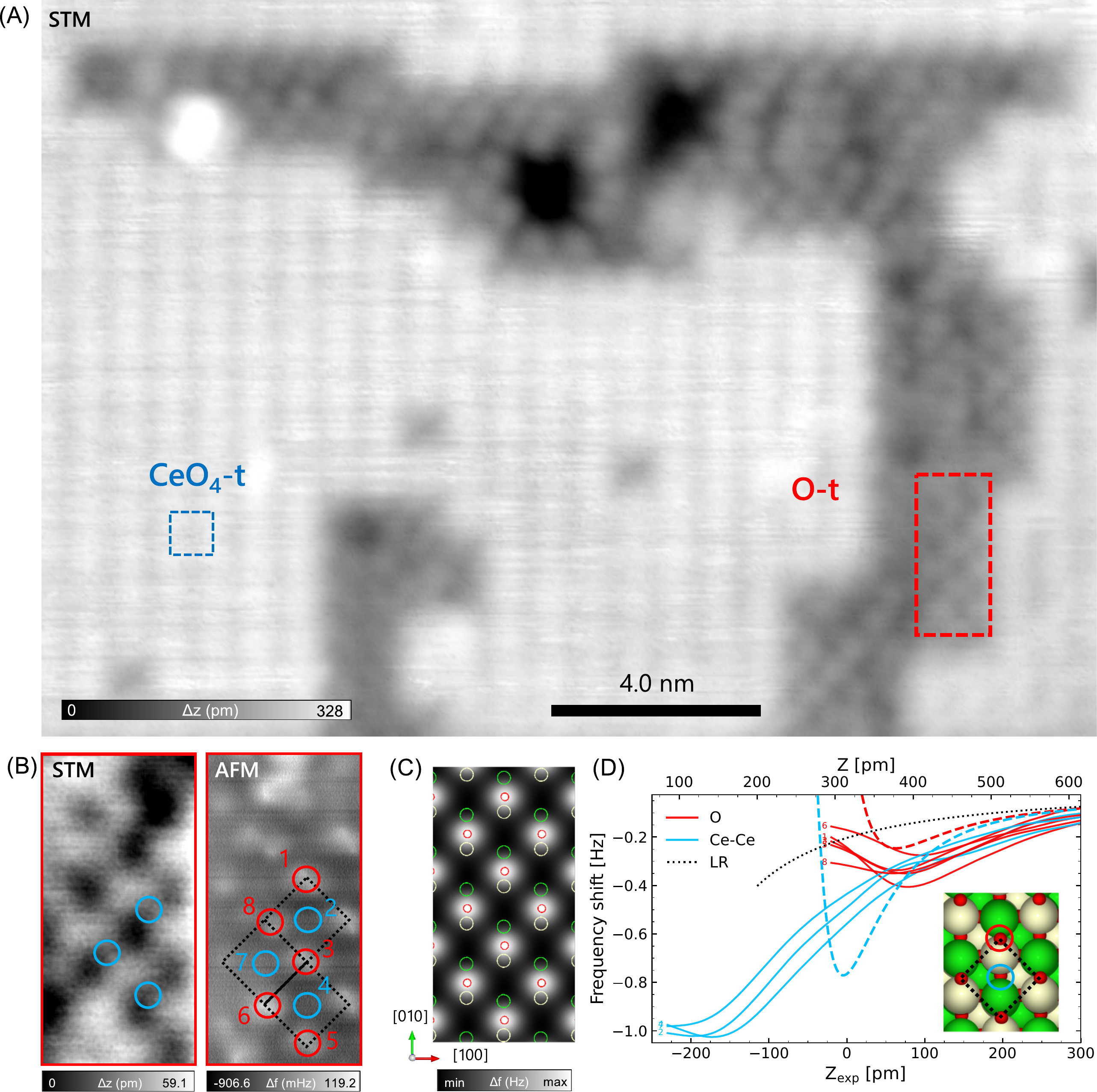}
\caption{%
\textbf{Structure and site-specific force spectroscopy of the O-t(100).}
(\textbf{A}) Topographic STM image showing patches of the O-t(100) reconstruction, which are located $\sim$185~pm below a terrace decorated with the CeO$_4$-t(100) surface (Fig.\ref{Fig1}D).
(\textbf{B}) Topographic STM and constant-height AFM images of the marked region in (A). These
images were not acquired simultaneously. Red and blue circles denote top-layer O and Ce bridge sites, respectively, where force spectroscopy measurements were performed.
(\textbf{C}) FDBM simulation of a constant-height $\Delta f$ map of the reduced O-t(100) model displayed as inset in (D) and obtained at a probe-surface separation of 350~pm from the top oxygen layer. 
It shows bright maxima over the top oxygen atoms and attractive dark contrast at the lower-lying Ce atoms.
(\textbf{D}) Experimental $\Delta f(Z_{exp})$ (solid) and DFT-based simulated $\Delta f(z)$ (dashed) curves for the sites marked in (B) and the inset model, respectively. 
The structural model %
corresponds to the O-t(100)-mix model, one of the two reduced O-t(100) surfaces considered (see Figure~S3D).
The black dotted line represents the long range (LR) vdW contribution of the mesoscopic part of the probe (see Materials and Methods).
Acquisition parameters: STM, $V=2.7$~V, $I=2$~pA; AFM, $V=100$~mV, $A=60$~pm, $f_0=992175$~Hz).
}
\label{Fig5}
\end{figure}

As in the case of CeO$_2$(100) islands grown on Ru(0001)~\cite{Niklas}, our films also display patches of the O-t(100) reconstruction (Fig.~\ref{Fig1}C~and~F).
Figure~\ref{Fig5}A shows a topographic STM image covering the second and third terraces along the profile in Fig.~\ref{Fig1}D.
A considerable portion of the image displays the CeO$_4$-t(100) surface, whereas areas of the third terrace expose the O-t(100) reconstruction.
In empty-states STM images ($V>0$), domains of O-t(100) are identified through their characteristic c(2$\times$2) periodicity, with a unit cell rotated by $45^\circ$ with respect to the CeO$_4$-t(100) one (Fig.~\ref{Fig1}C). 
An averaged nearest-neighbour distance between protrusions of~555~pm (see Figure~S3D)
points toward a reduced O-t(100) surface, with a 25\% of O$^{2-}$ ions with respect to the bulk-truncated CeO$_2$(100) surface~\cite{Niklas}.
As we show in the next section and at variance with the CeO$_4$-t(100) case, in a reduced O-t(100) surface the empty-states STM images do not directly map the Ce$^{4+}$ sites.
Thus, we rely once again on constant-height AFM imaging and site-specific force spectroscopy to determine the position, chemical termination, and local reactivity of the atomic sites.
Similarly to the topographic STM images, at the probe-surface separation used to acquire the AFM image in Fig.~\ref{Fig5}B, the contrast is dominated by bright spots in a c(2$\times$2) arrangement. 
By comparison with the reduced O-t(100) structural models (Fig.~\ref{Fig1}F, and Figures S3D and S4)
and in view of site-dependent force spectroscopy measurements (Fig.~\ref{Fig5}D), this bright signal seems to correspond to the onset of a repulsive interaction with the oxygen atoms at the top of the surface, while the interaction with the lower-lying Ce atoms is attractive in nature, giving rise to a dark contrast in the AFM image.

This hypothesis is confirmed by first-principles simulations, in which two reduced O-t(100) models differing in the distribution of the Ce$^{3+}$ ions between the top and second Ce layers were considered (see Figures~S3D~and~S4).
Since both models expose the same top oxygen layer, they are expected to exhibit essentially the same AFM contrast, while their different Ce$^{3+}$ distributions primarily affect the STM contrast, as discussed below.
The experimental $\Delta f(Z_{exp})$  spectroscopy acquired over the bright AFM spots in Fig.~\ref{Fig5}B is reproduced by DFT-based simulated $\Delta f(z)$ curves over the oxygen sites of a reduced O-t(100) model (Fig.~\ref{Fig5}D); calculated curves that also include the long-range vdW contribution from the mesoscopic part of the probe (see Materials and Methods).
Alignment of experimental and calculated curves at the minima produces an imaging height of $Z \approx 325$~pm with respect to the plane defined by the top oxygen atoms.
The simulated curves over the Ce--Ce bridge positions probed in the experiments (Figs.~\ref{Fig5}B~and~D) display minima of comparable magnitude to the experimental counterparts, but are shifted by $\sim$125~pm closer to the surface.
We find that both the FDBM-based AFM images and the calculated $\Delta f(z)$ curves  provide better agreement with the experimental data for a reduced O-t(100) surface than for the CeO$_4$-t(100) case. 
At the relevant imaging heights of the experiment, the FDBM force decomposition for AFM images of the O-t(100) (Figure~S7 and S8)
shows that the contrast is mainly governed by short-range Pauli repulsion from the top oxygen layer. 
The electrostatic interaction acts as a small modulation, while the vdW dispersion correction provides an almost flat attractive background.
This interaction balance differs from the CeO$_4$-t(100) case 
and explains why a rigid CO probe model provides a fair description for AFM simulations on the O-t(100) surface.
When the contrast is governed mainly by short-range Pauli repulsion, the force response is predominantly controlled by the localized charge density at the forefront part of the apex, in particular, by the lone-pair associated with the oxygen atom.
The rigid CO model captures this front-end density sufficiently well to reproduce the AFM contrast for the O-t(100) surface and corresponding $\Delta f(Z_{exp})$ curves.
In contrast, when electrostatics dominates (as for several sites of the CeO$_4$-t(100) reconstruction), the force response becomes more sensitive to the extended charge distribution and detailed structure of the Cu$_x$O apex, which the rigid CO model fails to capture in full  (see the discussion in the section S3.2).

\subsubsection{Origin of the STM contrast in the O-t(100) surface}

\begin{figure}[b!]
    \centering
    \includegraphics[width=1.0\linewidth]{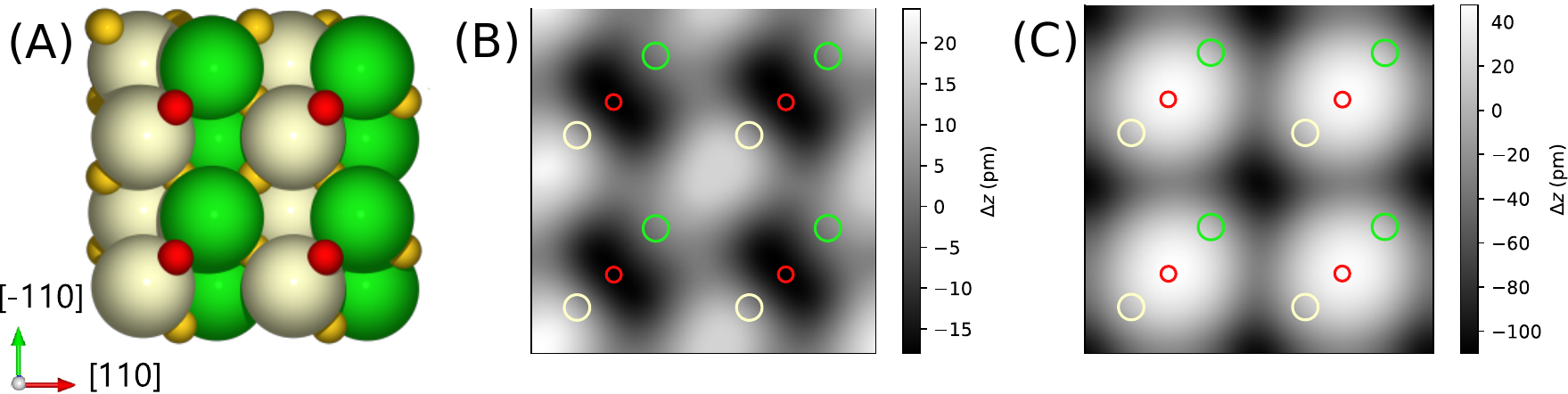}
\caption{%
\textbf{Simulated constant-current STM images of the reduced O-t(100) surface with Ce$^{3+}$ ions distributed across the first two Ce layers.}
(\textbf{A}) Structural model of the reduced O-t(100) termination with Ce$^{3+}$ ions distributed across the first two Ce layers.
(\textbf{B}, \textbf{C}) Simulated empty- and filled-state STM images, respectively.
The filled-state contrast is centered on the top-layer oxygen atoms, whereas the empty-state image displays asymmetric maxima associated with the vacuum decay of the unoccupied density from the top-layer Ce$^{4+}$ sites.
Isovalues correspond to the mean value in a plane 3~$\mathring{\mathrm{A}}$ above the top oxygen layer. The empty- and filled-state isovalues are $2.56\times10^{-4}$ and $3.57\times10^{-5}$~states~$\mathring{\mathrm{A}}^{-3}$, respectively.
}
\label{Fig6}
\end{figure}

Upon establishing that this second reconstruction explored in the experiments is oxygen terminated and has a reduced character, we investigated the possibility of using STM to gather information about the Ce$^{3+}$ ions that should populate near-surface layers.
To this end, we simulated constant-current STM images for two representative reduced O-t(100) models: 
one with Ce$^{3+}$ ions distributed between the first and second Ce layers (denoted O-t(100)-mix, Fig.~\ref{Fig6}A) and another with the Ce$^{3+}$ ions localized in the first Ce layer (O-t(100)-top, Figure~S13A). 
These two models are equally stable (see the discussion in Figures~S3~and~S4)
and produced, according to our simulations, almost identical AFM images and force spectroscopy curves.

For both models, the contrast in filled-state STM images (V$<$0) is mainly governed by O-derived occupied states (Figures~S17~and~S16)
and follows the positions of the top oxygen atoms (see Fig.~\ref{Fig6}C and Figure~S13C). 
In empty-state imaging (V$>$0), the situation is more subtle.  
In the O-t(100)-mix model, the empty-state projected density of states is dominated by unoccupied states from the Ce$^{4+}$ ions in the first Ce layer.
However, the corresponding integrated LDOS develops a nonspherical and laterally shifted lobe in the vacuum (Figure~S16),
so that the STM contrast maxima do not coincide directly with the atomic positions of the Ce$^{4+}$ ions, but they extend over Ce-Ce bridge site \textemdash position that should be occupied by an oxygen atom in the fully oxidized surface, see Figure~S3C\textemdash\space
towards the neighboring Ce$^{3+}$ ion (Fig.~\ref{Fig6}B).
 This result is consistent with the assignment of chemical species in the experimental STM image (Fig.~\ref{Fig5}B) provided by the AFM images and $\Delta f(Z)$ spectroscopy, and is also in agreement with displaced STM signals toward similar regions with missing surface oxygen atoms reported for the  reduced CeO$_2$(111) surface~\cite{NiliusVeronica}.

When the first Ce layer is completely reduced (fully populated by Ce$^{3+}$ ions, see the O-t(100)-top model in Figures~S3D~and~S13A), our simulations show a very similar STM contrast pattern (Figure~S13B),
but with current values that are approximately one order of magnitude smaller than for the O-t(100)-mix model (Fig.~\ref{Fig6}), as evidenced by the corresponding isosurface values stated in the corresponding figure caption. 
The integrated LDOS within the selected empty-state energy window is again dominated by the unoccupied Ce$^{4+}$ states.
However, these states are localized in deeper layers and therefore do not contribute significantly to the electronic density tunneling into the vacuum.
The simulated STM contrast thus arises from %
this small Ce$^{4+}$ contribution combined with evanescent tails of Ce$^{3+}$ and oxygen states
from different atomic layers.
The total of these contributions produces a pattern in which the intensity maxima are displaced from the uppermost oxygen atoms towards the midpoint between two neighboring Ce$^{3+}$ sites (Figure~S17).
As these Ce$^{3+}$ and oxygen states have only a very small weight within the empty-state LDOS, the resulting state density in the vacuum is extremely low, leading to small isosurface values \textemdash and therefore tunneling currents\textemdash\space obtained in the simulations.
Such currents are expected to be negligible at the probe-surface separations typically employed in STM experiments.
Our simulations show that both configurations reproduce the observed c(2$\times$2) periodicity in empty-states STM images, however, the current maxima provide neither a direct registry with the top-layer oxygen atoms nor the Ce sites (either Ce$^{3+}$ or Ce$^{4+}$) in the first Ce layer. 
This clarifies a limitation when trying to assign the O-t(100) atomic registry from the periodicity observed in empty-state STM images alone, as was done in previous STM-based works~\cite{Niklas}.
Instead, the apparent STM maxima are determined by the spatial decay of the integrated LDOS into the vacuum, which is displaced towards the oxygen surface vacancies \textemdash the Ce-Ce bridge position in Fig.~\ref{Fig5} where the top oxygen atoms of the fully oxidized O-t(100) surface should be located, %
compare Figures~S3C
and~S3D.

Notice that the isosurface value for the O-t(100)-top model (Figure~S13A) is of the same order of those found in our simulations for filled-states imaging in the CeO$_4$-t(100) reconstruction (Fig.~\ref{Fig3}C and F), where no contrast was observed in the STM experiments (Fig.~\ref{Fig2}D).
Consequently, we propose that the experimental STM contrast observed for the O-t(100) reconstruction (Fig.~\ref{Fig5}B) can only be explained by the O-t(100)-mix model, where the Ce$^{3+}$ ions are distributed between both the first and second Ce layers.

\subsection{Categorization of the X$_{p}$ feature observed at the CeO$_4$-t(100) surface}

In addition to the two-lobed AFM signal that captures the interaction of the probe with two oxygen atoms at adjacent CeO$_4$ pyramids (red circles and $\Delta f(z)$ curves in Figs.~\ref{Fig4}), the image in Fig.~\ref{Fig4}A shows several single bright spots (labeled X$_{p}$ in Fig.~\ref{Fig4}B) that are nearly oval and share similar surface location.
These X$_{p}$ features are positioned asymmetrically between the Ce atoms (see Figure~S2),
and present markedly different $\Delta f(Z)$ curves (purple in Fig.~\ref{Fig4}C): their minima are shallower, occur $\sim$50~pm farther from the surface and reach roughly half the maximum attractive interaction obtained at the sites between the oxygen atoms of adjacent pyramids. 
Beyond the minimum, these curves bend back towards more attractive values; a characteristic signature of local probe-surface relaxations, as previously reported for reduced oxide surfaces~\cite{oscar_manuel_ceria_111_water} and for AFM-based vertical manipulations~\cite{Sugimoto08-Science-VerticalInterchangeManipulation}.
The presence of these X$_{p}$ features appears to influence the dispersion of the $\Delta f(Z)$ curves measured at adjacent Ce sites (the minima of the curves measured over the Ce atoms labeled 3-9 and 13-15 show greater variability), but no single monotonic correlation is observed. 
We therefore treat these X$_{p}$ entities as local perturbations of the pristine CeO$_4$-t(100) lattice that modify the site-dependent interaction with the AFM probe.

The single bright AFM contrast, the asymmetric position, the early onset of the force response, and more importantly, the bending of the curve suggest that the X$_p$ feature could be associated with an adsorbate species. In view of
the small attractive 
minimum and the early onset of repulsive interactions, this adsorbate is likely to involve an oxygen atom protruding outward from the surface.
We nevertheless examined whether these characteristics could arise from rearrangements of intrinsic surface oxygen atoms by testing several models in which oxygen atoms were displaced from the CeO$_4$ pyramids to nearby sites.
After relaxation, however, these structures either converged to less stable configurations or their calculated AFM images failed to reproduce the single-lobed and strongly repulsive features observed.
Purely intrinsic rearrangements of the pristine lattice are therefore unlikely to account for the X$_p$ entities.

\begin{figure}
\centering
\includegraphics[width=1.0\linewidth]{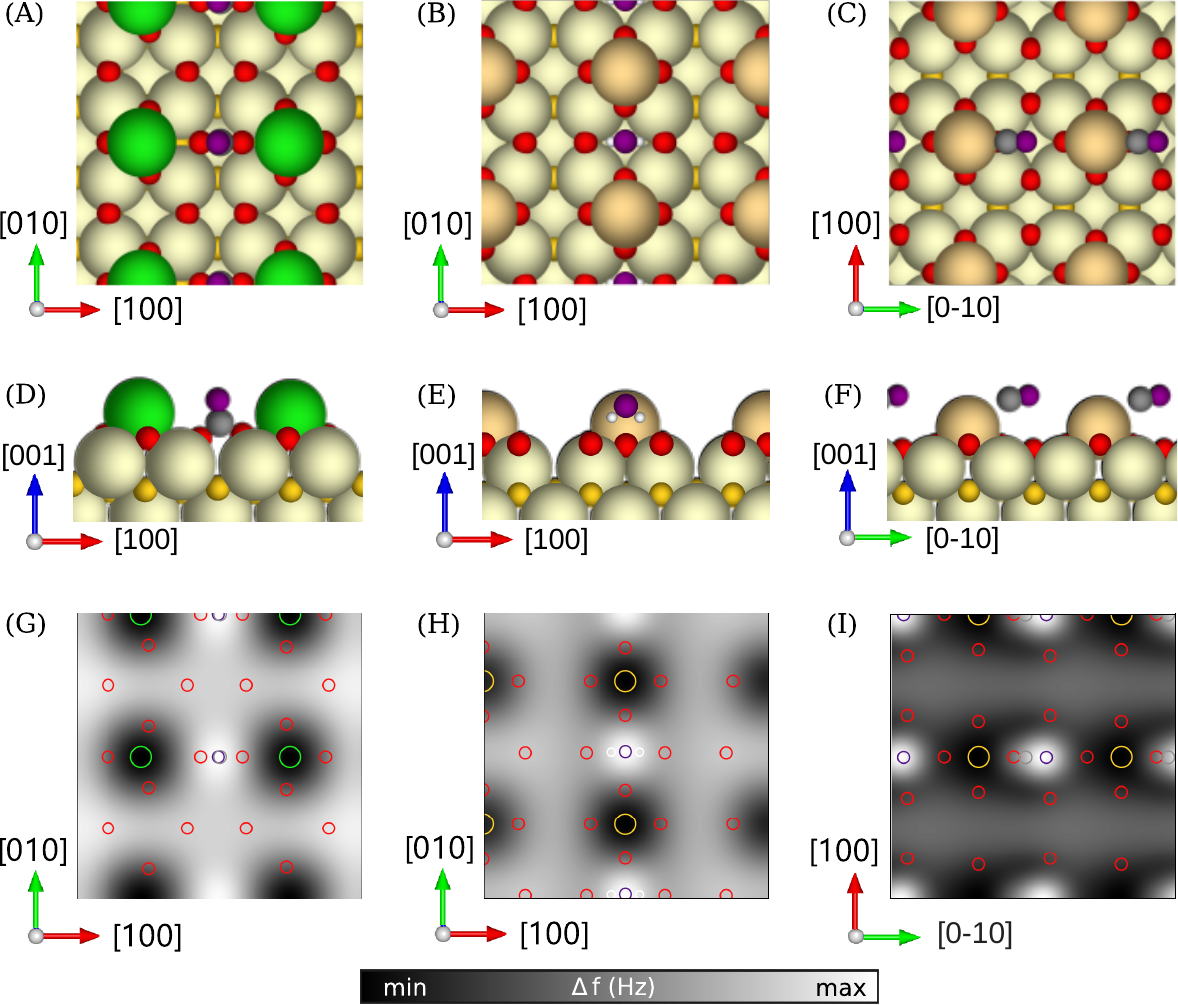}
\caption{%
\textbf{Adsorbate models reproducing the X$_p$ contrast on CeO$_4$-t(100).}
(\textbf{A--C}) Top and (\textbf{D--F}) side views of DFT-optimized [CO$_3$]$^{2-}$, H$_2$O, and CO adsorption geometries. 
(\textbf{G--I}) Corresponding FDBM constant-height $\Delta f$ maps at $Z=500$~pm.
All three models generate a single bright contrast between the CeO$_4$ pyramids observed in the experiments for the X$_p$ entities.
Crystallographic directions are indicated in each panel.
A different orientation in (C), (F) and (I) is selected for a clear visualization of the adsorbate. 
Colors: Ce$^{3+}$ (green), surface O (red and yellow), Ce$^{4+}$ (orange and beige), C (gray), and H (white).
}

\label{Fig7}
\end{figure}

According to our hypothesis, we considered molecular species containing oxygen atoms that can be present in the residual vacuum of the UHV system, in particular H$_2$O and CO~\cite{Eren2021_H2OandCO}.
For CO, both the intact molecule and a carbonate-like [CO$_3$]$^{2-}$ species were considered as possible adsorbates, the latter formed by coordination with two oxygen atoms of adjacent CeO$_4$ pyramids~\cite{Albrecht2014May}.
DFT calculations were performed to optimize the adsorption geometries of these species on the CeO$_4$-t(100) surface (see Materials and Methods), positioning each adsorbate in a surface location similar to the one detected in the experiments, with an oxygen atom pointing towards the vacuum.
The top and side views for the relaxed structures for [CO$_3$]$^{2-}$, H$_2$O, and CO are shown in Figs.~\ref{Fig7}A~to~F, together with the corresponding simulated constant-height AFM images (Figs.~\ref{Fig7}G~to~I).
In all cases, the simulations reproduce the characteristic bright, oval-shaped AFM maximum centered between the Ce atoms.
The selected adsorption geometries reproduce the experimentally observed X$_p$ contrast, although they do not necessarily correspond to the global minimum configurations.
Note that, with the exception of CO, these configurations are not the lowest-energy adsorption geometries (see Figure~S5):
they were intentionally selected to reproduce the experimentally observed AFM contrast.

For [CO$_3$]$^{2-}$, a flat structure between the pyramids is 0.72~eV more stable than the vertical adsorption shown in Fig.~\ref{Fig7}A, which just needs to overcome a 0.06 eV barrier to become flat. 
DFT calculations also show an even more stable flat carbonate (0.74~eV) formed with O atoms that do not belong to any pyramid (Fig.~S5);
structure only possible at the fully oxidized surface.
For H$_2$O, a structure similar to the one presented in Fig.~\ref{Fig7}B but with the molecule orienting its O atom towards the Ce atom of the pyramid is 0.13~eV more stable.
These results indicate that adsorption of common residual gases could account for the X$_{p}$ features observed at the CeO$_4$-t(100) surface, leaving the precise chemical identity of the adsorbate open for further investigation.

\section{Conclusions}

By combining atomic-resolution STM and AFM imaging with site-specific force spectroscopy and first-principles simulations, we explore the possibility of correlating local reactivity and reduction state to the atomic structure of a chemically active oxide surface.
Our results support earlier experimental and theoretical predictions for the Ce terminated (2$\times$2) \textemdash named CeO$_4$-t(100)\textemdash\space and oxygen-terminated c(2$\times$2) \textemdash in short O-t(100)\textemdash\space reconstructions of the CeO$_2$(100) surface~\cite{Niklas,STETSOVYCH2013766,Capdevila-Cortada2017Mar}, and add direct experimental evidence on their structure and composition that was lacking in previous works. 
Furthermore, our DFT study of the electronic structure and site-specific AFM interaction of the atomic species populating these surfaces reveals important limitations of the STM for the characterization of the atomic structure and unveils the nature of the chemical interaction of the surface atoms with the oxygen-terminated probe used in the experiments, which in turn exposes a limitation when modeling the AFM probe.

Beyond the periodicity accessible by STM alone in this and previous works, AFM and force spectroscopy enable direct visualization of the atomic structure and categorization of the atomic species populating the surface: Ce and O sites together with alien species (named X$_p$) are observed for the CeO$_4$-t(100), and oxygen atoms are confirmed for the O-t(100).

Our simulations show that STM maxima in filled- and empty-state images cannot be trivially assigned to the position of one of the two chemical species.
In fact, the STM contrast results from a subtle competition between the spatial decay of the integrated local density of states into the vacuum and the three-dimensional arrangement of the surface and subsurface atoms.
This complex interplay  is not captured by the commonly used projection of the electronic states on short-range orbitals localized on the Ce and O sites.
As a consequence, for the  CeO$_4$-t(100) surface, STM contrast does not provide a robust fingerprint of the local reduction state, as oxidized and reduced terminations produce very similar STM images.
However, in the case of  the O-t(100) reconstruction,  STM is able to reveal the reduced state and the presence of a mixed distribution of Ce$^{3+}$ and  Ce$^{4+}$ sites on the first underlying Ce layer.

Site-specific force spectroscopy in the form of $\Delta f(z)$ curves and first-principles-based simulated AFM images reveal that for the O-t(100) surface, the AFM contrast is mainly governed by short-range Pauli repulsion from the top oxygen layer, which explains the good agreement obtained between simulations and experiments using a rigid CO to model the probe.
For the CeO$_4$-t(100) case, in contrast, the electrostatic interaction plays a major role, making the force response more sensitive to the extended charge distribution and detailed structure of the experimental Cu$_x$O-terminated apex.
This characteristic limits the use of simple probe models to describe experimental results on polar, strongly corrugated oxide surfaces, particularly for sites located below the topmost atomic layer.

We deployed a combination of experiments and first-principles simulations to try to categorize the nature of an unknown alien species observed as a defect at the surface of CeO$_4$-t(100).
The characteristic response in AFM images and spectroscopic curves over this defect points towards an oxygen-containing gas molecule from the residual vacuum adsorbed on the surface.
Selected local-minimum configurations of CO, H$_2$O, and carbonate-like [CO$_3$]$^{2-}$ reproduce the main experimental AFM signatures, although the precise chemical identity of this alien species remains open.

This work expands the atomic-scale understanding of the CeO$_{2}$(100) surface and highlights the strength of combining STM and AFM imaging with site-specific force spectroscopy, supported by first-principles calculations, to investigate atomic structure, local chemical interactions, and adsorbate-induced perturbations on wide-band-gap metal oxide surfaces.

\section*{Data Availability Statement}

The data supporting the findings of this study are available in the Zenodo repository at
\href{https://doi.org/10.5281/zenodo.21628218}{10.5281/zenodo.21628218}.
The repository contains \texttt{Structures.tar.gz}, with the VASP structural files for the models reported in this work; \verb+ DFT_force_data.zip + and \verb+ DFT_fs_data.zip +, with the calculated force and frequency-shift data; and \verb+ experimental_fs_data.zip +, with the experimental frequency-shift data used to reproduce the corresponding figures.
A detailed description of the structural data is provided in \texttt{Structures\_README.txt}.

\section*{Supporting Information}

\noindent
 Additional experimental data analysis to support statements in the main text is included in Figures~S1~and~S2; DFT optimized atomic structures for oxidized and reduced models of the CeO$_4$-t(100) and O-t(100) reconstructions, as well as for candidates to describe the X$_p$ perturbation are presented in Figures~S3~and~S4, and Figure~S5, respectively; the decomposition of forces based on the FDBM formalism for the CeO$_4$-t(100) and O-t(100) reconstructions, and for the candidates to describe the X$_p$ perturbation are included in Figure~S6, Figures~S7~and~S8, and Figures~S9~to~S11, respectively; geometries of several Cu$_2$O clusters considered as alternative probe models are presented in Figure~S12; simulated constant-current STM images of the reduced O-t(100) surface with Ce$^{3+}$ ions confined to the top subsurface Ce layer are depicted in Figure~S13; projected density of states and STM isosurfaces for the oxidized and reduced CeO$_4$-t(100) reconstruction, and for two distributions of Ce$^{3+}$ ions for the O-t(100) surfaces are presented in Figures~S14~and~S15, and Figures~S16~and~S17, respectively (PDF)

\section*{Author Contributions}

KK and MGL contributed equally to this work. KK and OC carried out the experiments, analyzed the experimental data, and prepared the original draft of the manuscript under the supervision of MA and SK. HSA provided technical assistance in sharpening the AFM probes. EFV built and optimized the atomic models under the supervision of MVGP. MGL and PP performed the electronic-structure analysis together with the AFM and STM simulations based on these optimized models under the supervision of RP. KK, OC, MGL, PP, EFV, MVGP, and RP discussed the results, contributed to the interpretation of the data, and revised the manuscript. All authors have read and approved the final version of the manuscript.

\begin{acknowledgement}
The experiments were carried out at NIMS, and the work was supported by NIMS grants AG2030 (OC, SK) and AM2100 (KK, MA, OC); by several Grants-in-Aid for Scientific Research from Ministry of Education, Culture, Sports, Science and Technology (MEXT, Japan): 19H05789 (OC, MA), 26H02244 (MA), 21K18876, 22H00285 (OC, SK), 23KJ1516 (KK), and 24K01350 (OC); by the Spanish Ministry of Science, Innovation and Universities (MCIU) through projects PID2023--149150OB--I00, PRE2021-098697, PID2021-128915NB-I00, and PID2024-162810NB-I00, and by the ``Mar\'{\i}a de Maeztu'' Programme for Units of Excellence in R\&D (CEX2023--001316--M). E.F.V.\ acknowledges support from the  the Margarita Salas postdoctoral fellowship (Spanish MIU and European Union NextGenerationEU) and the Momentum programme (MMT24-ICP-01: This research work was funded by the European Commission - NextGenerationEU, through Momentum CSIC Programme: Develop Your Digital Talent). The authors acknowledge the Spanish Supercomputing Network (RES) for computational resources at the MareNostrum Supercomputer (BSC, Barcelona). KK also acknowledges support from the Program for Leading Graduate Schools, ``Interactive Materials Science Cadet Program.'' The authors thank Dr.\ Jacek Goniakowski and Dr.\ Niklas Nilius for stimulating discussions on the atomic structure of the CeO$_2$(100) surface reconstructions and for their valuable comments on the results presented in this work.
E. F. V. Staff hired under the Generation D initiative, promoted by Red.es, an organisation attached to the Ministry for Digital Transformation and the Civil Service, for the attraction and retention of talent through grants and training contracts, financed by the Recovery, Transformation and Resilience Plan through the European Union's Next Generation funds.
\end{acknowledgement}

\bibliography{Updated_bib, additional}

\end{document}


\maketitle

\newpage

\section{Experimental images of surface defects and adsorbates}

This section provides additional experimental data supporting several statements of the main text.

\begin{figure}
\centering
\includegraphics[width=0.9\textwidth]{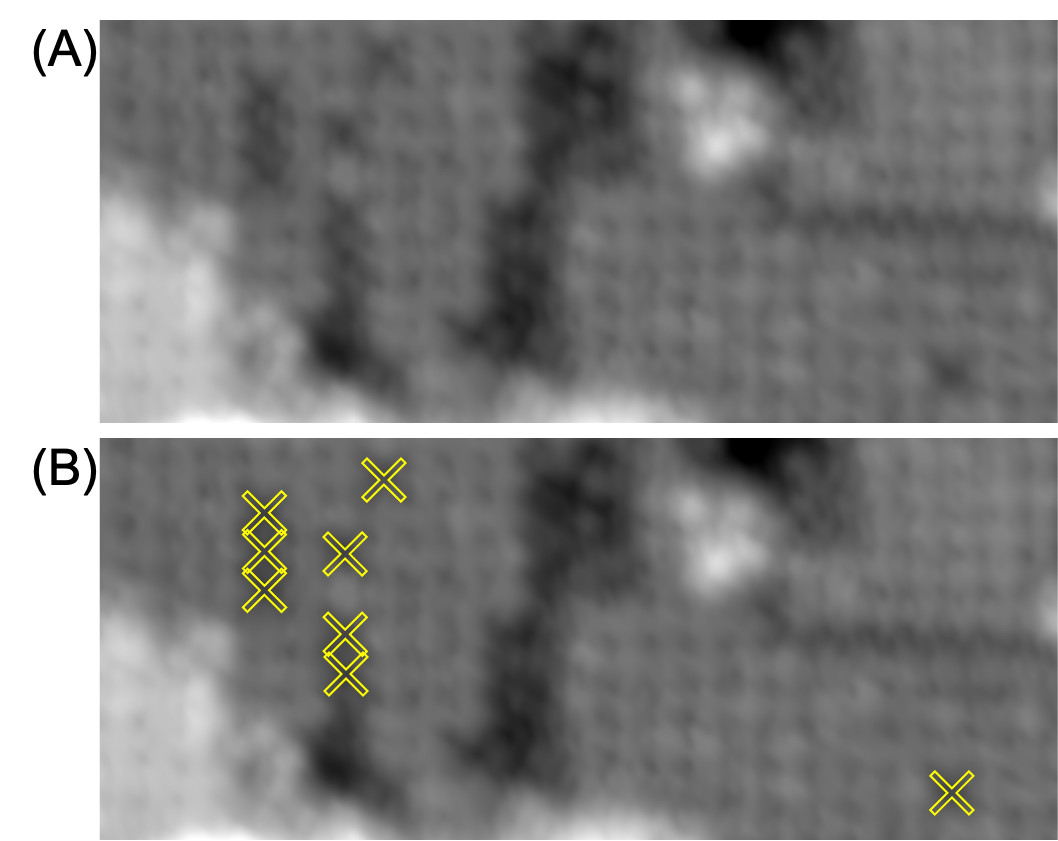}
\caption{
\textbf{Individual and linear arrays of Ce vacancies at the CeO$_4$-t(111) surface.}
(A) Close-up of the region in Fig.~\ref{Fig2}A showing individual and linear arrays of vacancies at Ce sites.
(B) Same STM image highlighting the position of each Ce vacancy.
Image size (18.5$\times$7.8)~nm$^2$. Imaging conditions as in Fig.~\ref{Fig2}.
}
\label{sfig:SI_linear_defect}
\end{figure}

\begin{figure}
\centering
\includegraphics[width=0.9\textwidth]{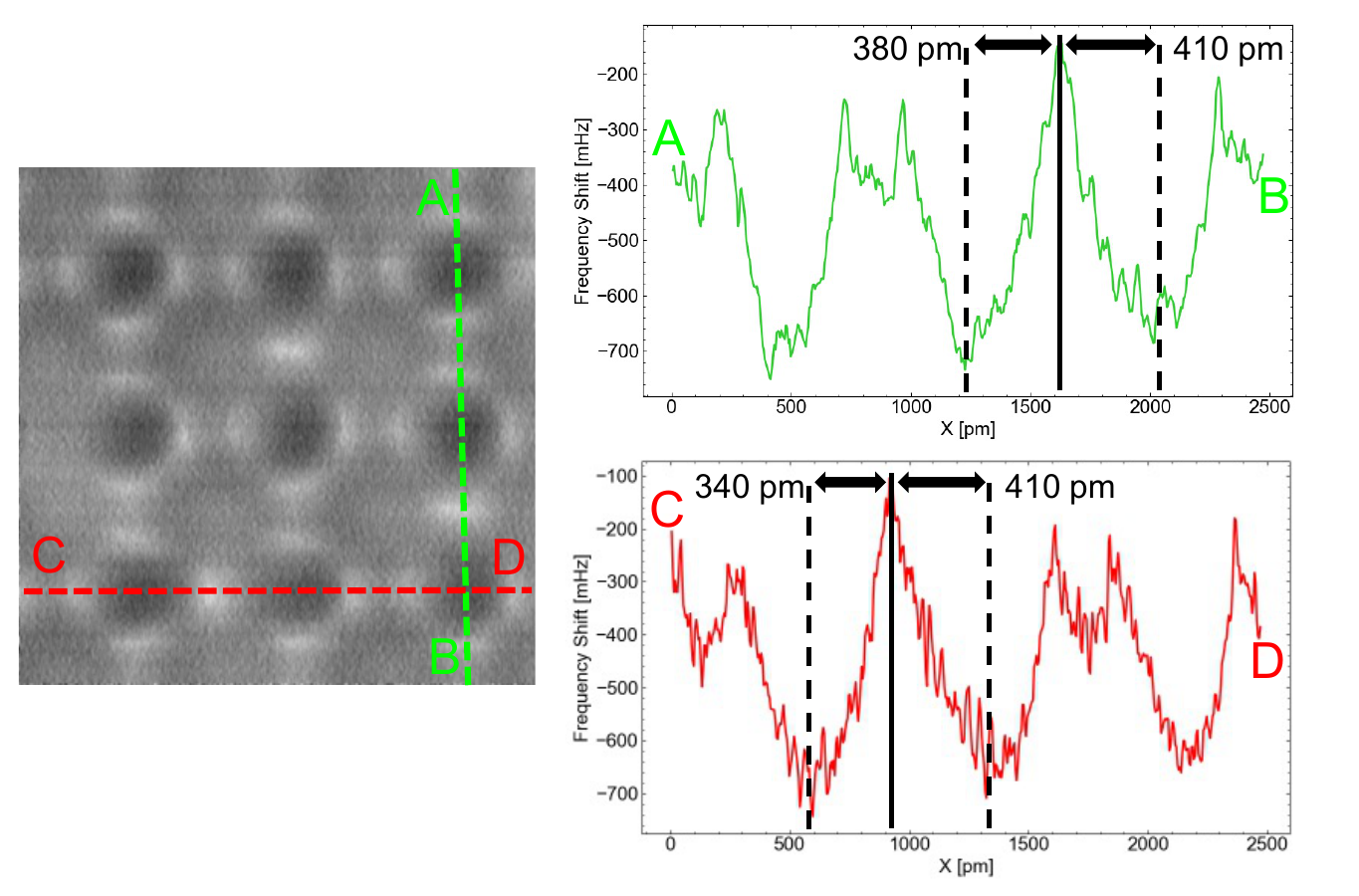} 
\caption{
\textbf{Asymmetric configuration of the X$_{p}$ feature.}
Left: constant-height AFM image of the CeO$_\mathrm{4}$-t(100) surface with two lines \textemdash A-B (in green) and C-D (in red)\textemdash\space crossing X$_{p}$ sites.
Right: corresponding $\Delta f$ line profiles. Black dashed lines mark the center of the Ce-site and the solid line the position of maximum contrast at the X$_{p}$ site. The distances from X$_{p}$ to the two neighboring Ce sites are unequal, e.g., 380~pm vs 410~pm along A-B and 340~pm vs 410~pm along C-D, demonstrating an off-center placement of the X$_{p}$ feature between adjacent Ce atoms. Imaging conditions are the same as in Fig.~\ref{Fig4}A.
}
\label{sfig:SI_Ostar_lineprofile} 
\end{figure}

\clearpage

\section{DFT structures for oxidized and reduced terminations and molecular adsorption}

This section gathers additional DFT-relaxed structures discussed in the main text, including oxidized and reduced surface models and adsorbate geometries considered to rationalize the X$_p$ feature.

\begin{figure*}[b!]
\begin{center}
\includegraphics[width=0.84\textwidth]{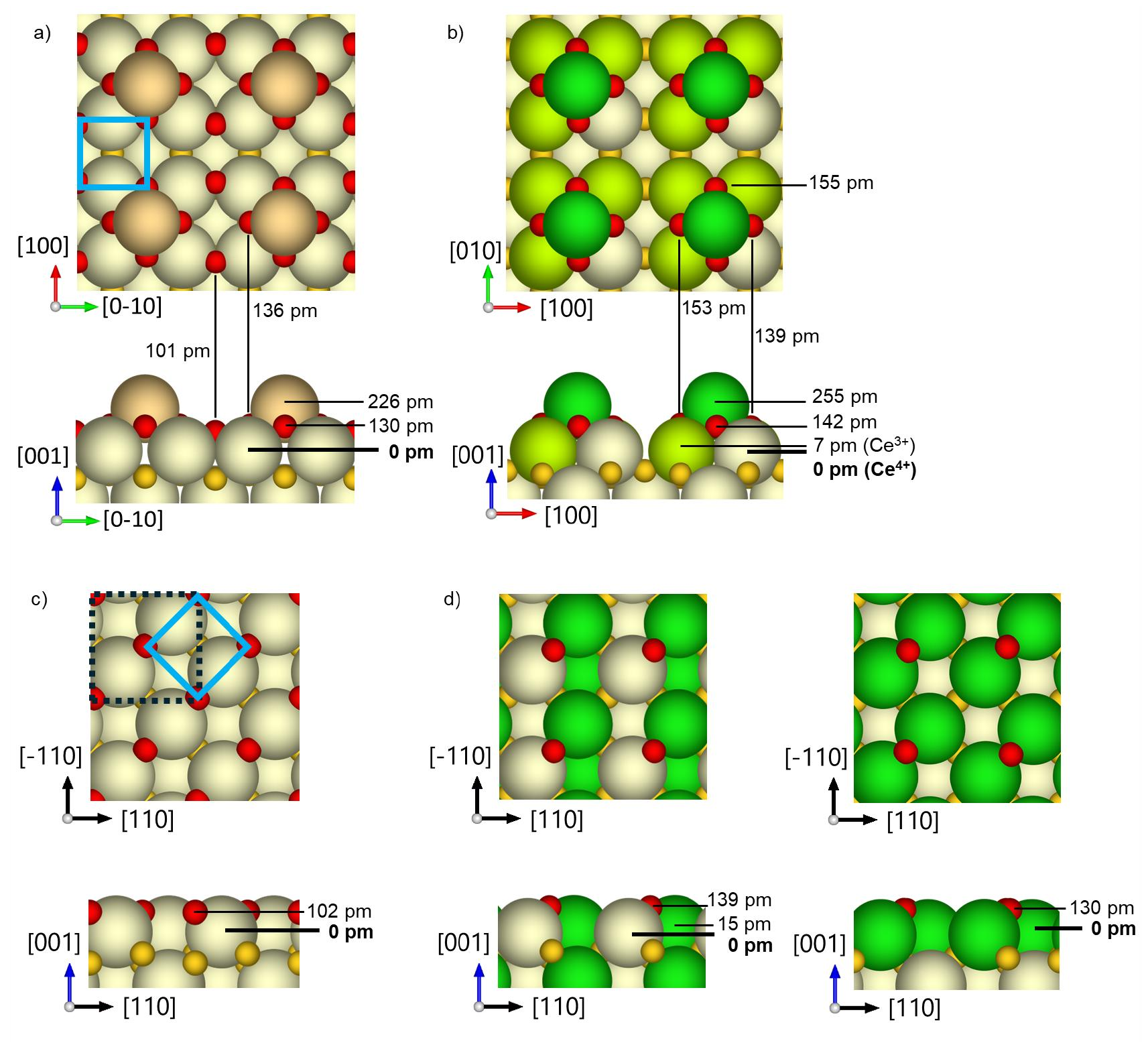}
\caption{
{\bf DFT-optimized structures for the oxidized (a) and the reduced (b) CeO$_\mathrm{4}$-t(100) surfaces, and for the oxidized (c) and the reduced (d) O-t(100) surfaces.} For (d), two arrangements for the Ce$^{3+}$ ions are considered, one with all Ce$^{3+}$ at the top layer (O-t(100)-top) and another with the Ce$^{3+}$ evenly distributed between the top and second layers (O-t(100)-mix), see also Fig.~\ref{sfig:SI_TopvsMixed}.The two configurations are essentially isoenergetic, differing by only 0.03 eV. Relative vertical $\Delta z$ distances between Ce and O atoms are indicated in pm. Surface O atoms in red (subsurface O atoms in yellow), Ce$^{4+}$ in sand (pyramid Ce$^{4+}$ ions in beige), Ce$^{3+}$ in light green (pyramid Ce$^{3+}$ ions in green). The primitive cells for both surfaces are marked in blue solid lines, together with the c(2$\times$2) unit cell in dashed black lines.}
\label{sfig:SI_DFT_Niklas}
\end{center}
\end{figure*}

\begin{figure*}[t!]
\begin{center}
\includegraphics[width=0.99\textwidth]{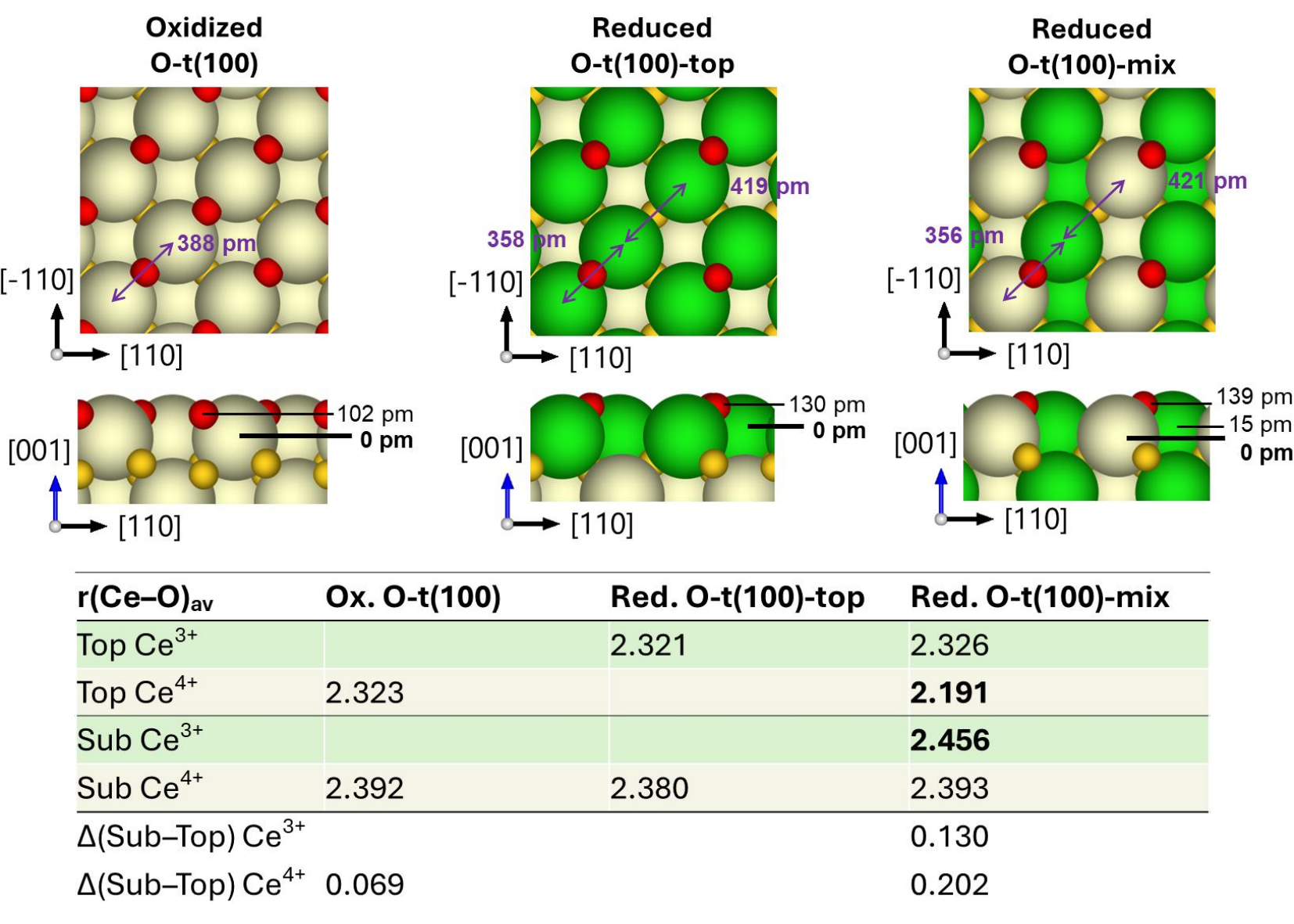}
\caption{
{\bf DFT-optimized structures for the oxidized and reduced O-t(100) surface.} Ce-Ce distances indicated in purple (top view), and relative vertical $\Delta z$ Ce-O distances in black (side view) are indicated in pm. Average Ce-O bond distances in the table (r(Ce-O)$_{av}$) in $\mathring{\text{A}}$. Surface O atoms in red (subsurface O atoms in yellow), Ce$^{4+}$ ions in sand and Ce$^{3+}$ ions in green.}
\label{sfig:SI_TopvsMixed}
\end{center}
\end{figure*}

For the reduced surface, we considered two arrangements of the Ce$^{3+}$ ions: one with all Ce$^{3+}$ ions located in the top layer (O-t(100)-top) and another with the Ce$^{3+}$ ions evenly distributed between the top and second layers (O-t(100)-mix), see Figs~\ref{sfig:SI_DFT_Niklas}~and~\ref{sfig:SI_TopvsMixed}. The two configurations are essentially isoenergetic, differing by only 0.03 eV. Based on the larger ionic radius of Ce$^{3+}$, one would expect the mixed configuration to be more stable, as distributing the Ce$^{3+}$ ions between the top and subsurface layers provides additional space for the expansion of their Ce–O coordination shell, whereas in the top configuration all Ce$^{3+}$ ions remain confined to the surface layer.~\cite{Murgida2013Jun,GandugliaPirovano2009Jan,NiliusVeronica}

The average Ce–O bond distances, r(Ce-O)$_{av}$, reveal the origin of the near degeneracy. In the O-t(100)-mixed configuration, the subsurface Ce$^{3+}$ ions indeed exhibit a larger average Ce–O distance (2.456~$\mathring{\mathrm{A}}$) than the top-layer Ce$^{3+}$ ions (2.326~$\mathring{\mathrm{A}}$). However, this gain is accompanied by a pronounced compression of the top-layer Ce$^{4+}$ ions, whose average Ce–O distance decreases to 2.191~$\mathring{\mathrm{A}}$, compared with 2.393~$\mathring{\mathrm{A}}$ for subsurface Ce$^{4+}$ ions and 2.323~$\mathring{\mathrm{A}}$ for the corresponding Ce$^{4+}$ ions in the oxidized surface. The stabilization associated with the larger Ce–O distances around the subsurface Ce$^{3+}$ ions is therefore offset by the stronger compression of the neighboring Ce$^{4+}$ ions, explaining why the two configurations remain nearly degenerate. These competing bond-length distortions originate from the surface oxygen vacancies, which create an alternating pattern of expanded and compressed Ce–O coordination shells, accompanied by minor rearrangements of the remaining surface oxygen atoms.

\begin{figure*}[h!]
\begin{center}
\includegraphics[width=0.94\textwidth]{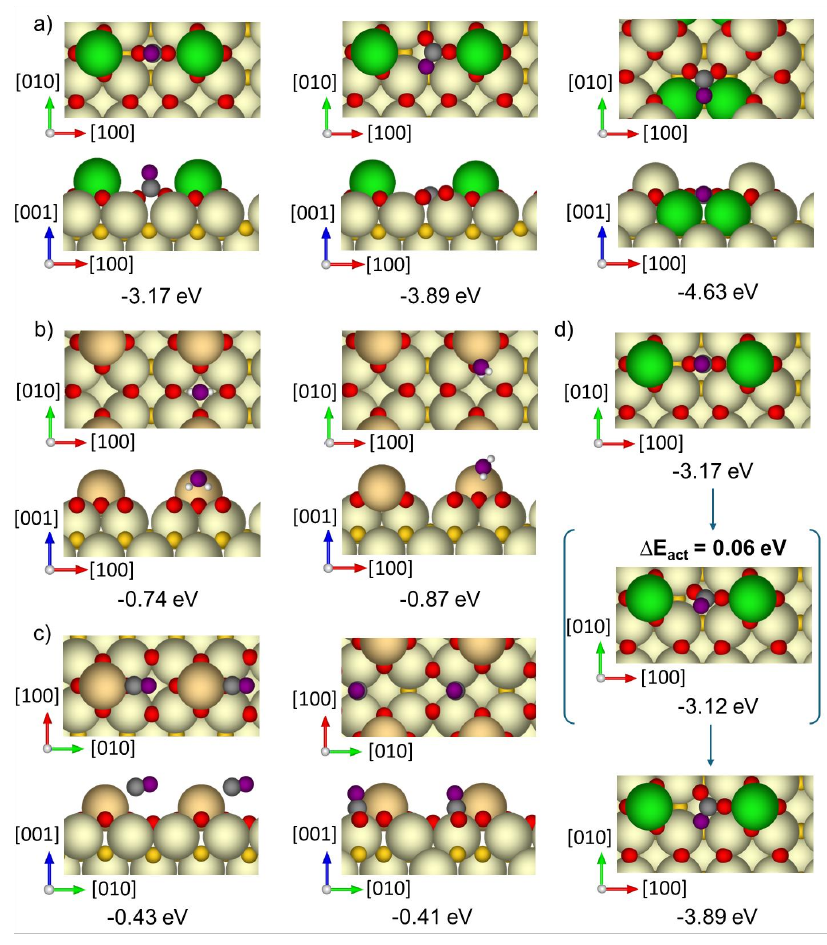}
\caption{
{\bf DFT-optimized (2$\times$4) structures for (a) CO adsorbed as [CO$_3$]$^{2-}$, (b) H$_2$O, (c) CO on the CeO$_\mathrm{4}$-t(100) surface, and (d) vertical to flat [CO$_3$]$^{2-}$ pathway.} Adsorption and activation energies are indicated in eV. Surface O atoms in red (subsurface O atoms in yellow), Ce$^{4+}$ in sand (pyramid Ce$^{4+}$ ions in beige), Ce$^{3+}$ in light green (pyramid Ce$^{3+}$ ions in green). 
For H$_2$O and [CO$_3$]$^{2-}$, here we also show structures that do not reproduce the experimental AFM contrast but that we found to be energetically more stable than the ones included in the main text, which potentially match the AFM experimental images. 
In the carbonate case, formation of [CO$_3$]$^{2-}$ from adsorbed CO and two surface O$^{2-}$ ions requires charge redistribution within the slab: two electrons remain in the carbonate moiety, while the other two localize on nearby Ce ions, giving rise to two Ce$^{3+}$ centers, consistent with the Bader-charge analysis.
}
\label{sfig:SI_DFT_adsorbates}
\end{center}
\end{figure*}

\clearpage

\section{AFM simulations: force decomposition and probe model} \label{secSI:afm_simulations}

This section presents additional AFM simulation data that support the interpretation of the site-dependent contrast discussed in the main text. Using the Full Density Based Model~\cite{Ellner2019} (FDBM), we decompose the  interaction into short-range, electrostatic, and van der Waals contributions for the pristine terminations and for the adsorbate models considered for the X$_p$ feature. We then discuss the limitations of the simplified probe model used to simulate the $\Delta f(z)$ signal for both constant-height AFM images and spectroscopic curves.

\subsection{FDBM force decomposition}

\subsubsection{Pristine surface terminations}

\begin{figure*}[h!]
\begin{center}
\includegraphics[width=0.99\textwidth]{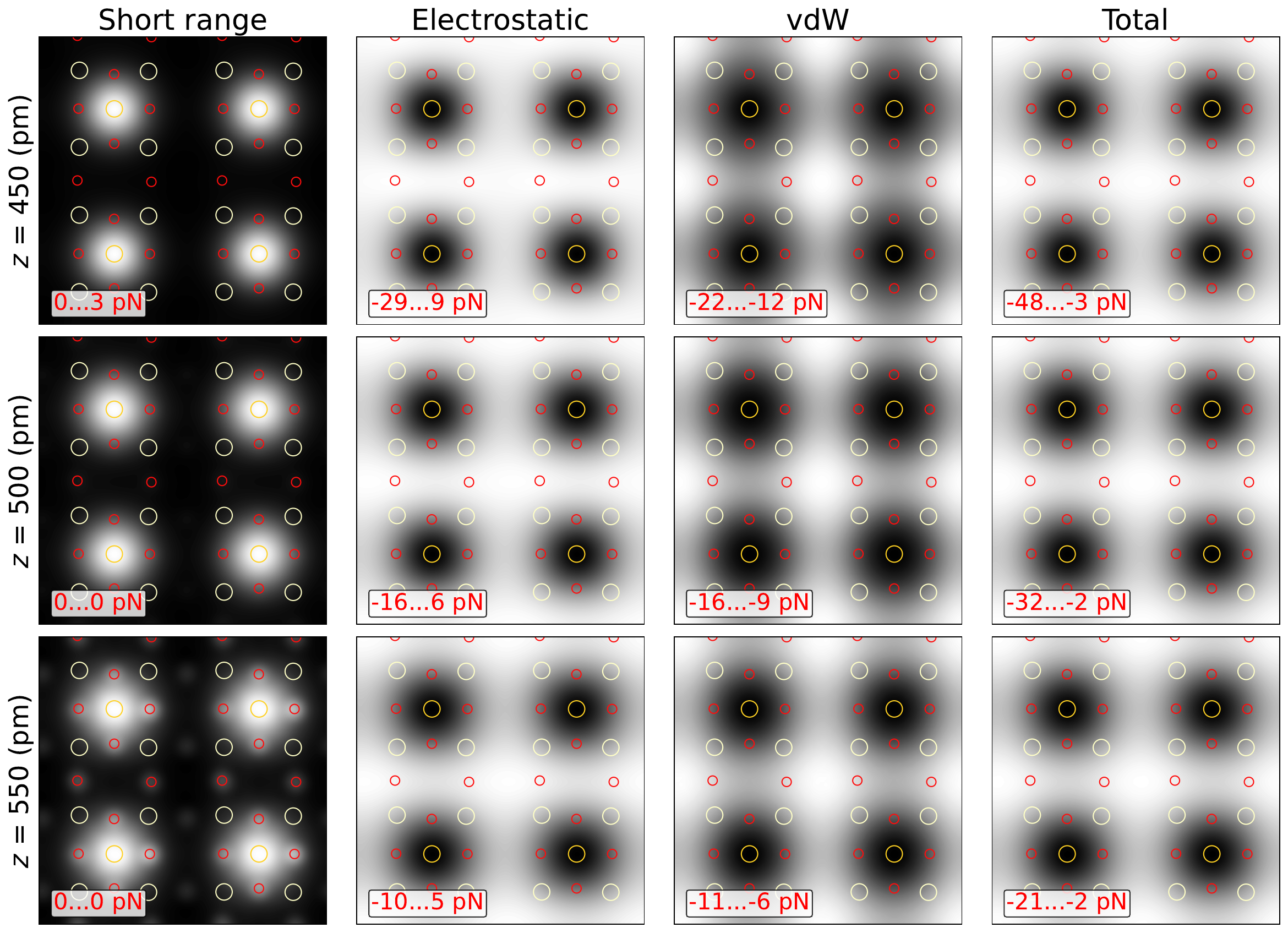}
\includegraphics[width=0.99\textwidth]{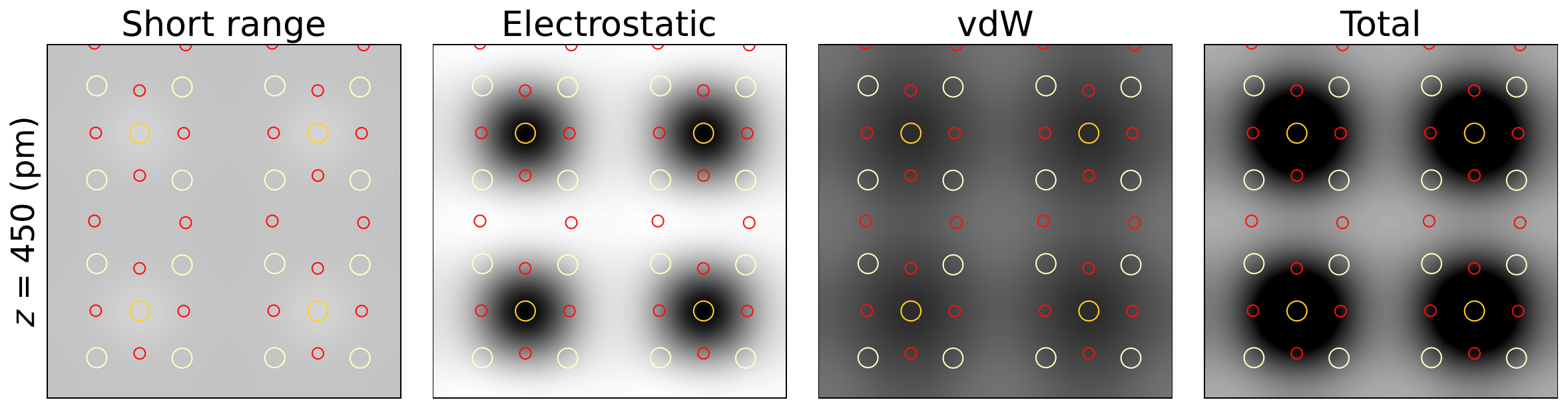}
\caption{
{\bf FDBM force decomposition for the CeO$_\mathrm{4}$-t(100) reconstruction.}
Decomposition of the AFM contrast obtained from Full Density Based Model (FDBM) calculations for the CeO$_\mathrm{4}$-t(100) reconstruction at three probe-surface separations ($z$) measured with respect to the position of the topmost Ce atoms of the pyramids.
The total AFM force (Total) is composed of short-range Pauli repulsion (SR), electrostatic (ES), and van der Waals (vdW) contributions. 
For the top three rows, each of the panels uses its own gray scale, with the minimum (black) and maximum (white) force values displayed as insets.
The panels at the bottom row represent the images for $z=450$~pm with the same absolute scale limits for all of them: (black, white) = (-48, 9)~pN.
These values correspond to the minimum (more attractive) and maximum (less attractive) force values found across the four panels; in this case, the minimum value (-48~pN) is reached in the Total force and the maximum (+9~pN) in the ES contribution.
This representation evidences that for this reconstruction, the electrostatic contribution dominates the AFM image, as it is shown by the resemblance between the contrast observed in the ES and Total panels. 
}
\label{sfig:SI_fdbm_2x2}
\end{center}
\end{figure*}

\begin{figure*}[h!]
\begin{center}
\includegraphics[width=0.99\textwidth]{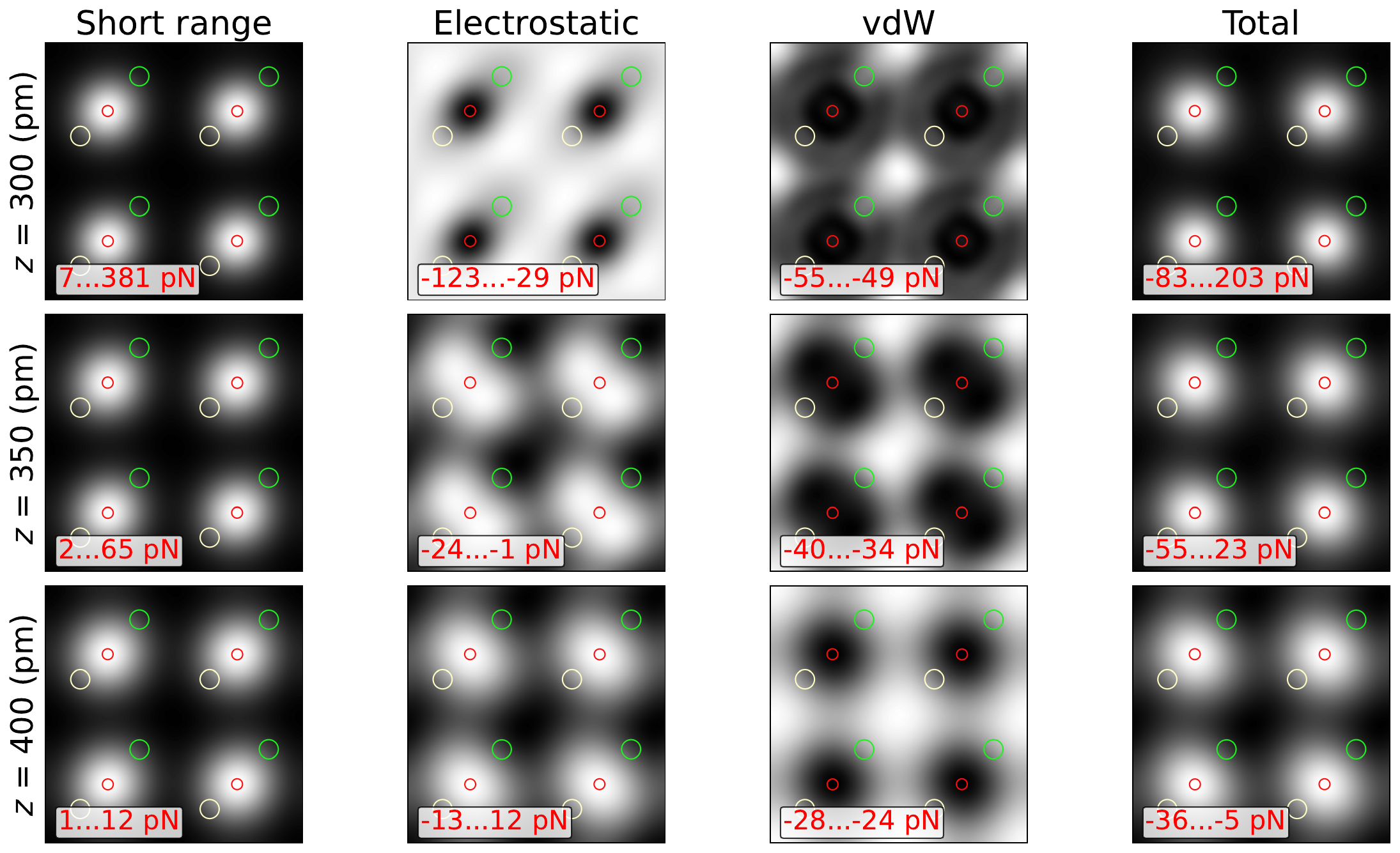}
\includegraphics[width=0.99\textwidth]{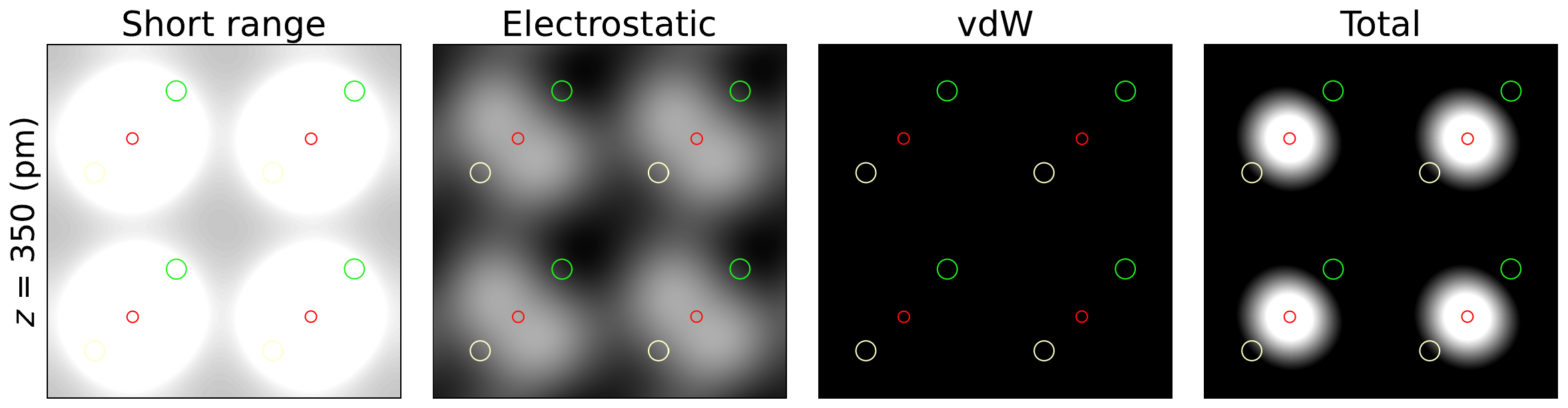}
\caption{
\textbf{FDBM force decomposition for the O-t(100) B2 reconstruction with Ce$^{3+}$ distributed across the first two Ce layers (O-t(100)-mix).}
The total force and its short-range (SR), electrostatic (ES), and van der Waals (vdW) components are shown at three probe--surface separations, $z$, referenced to the topmost surface O atoms.
The upper rows use independent grayscale ranges, whereas the bottom row shows all contributions at $z=350$~pm on a common scale of $(-25,10)$~pN.
As for the model with all Ce$^{3+}$ ions in the subsurface layer (Fig.~\ref{sfig:SI_fdbm_O_t_subsurf}), the AFM contrast is dominated by the short-range interaction with the upper O layer.
Electrostatics produces only a secondary modulation along the O rows, while the vdW contribution remains nearly uniform.
}
\label{sfig:SI_fdbm_O_t_2_layers}
\end{center}
\end{figure*}

\begin{figure*}[h!]
\begin{center}
\includegraphics[width=0.99\textwidth]{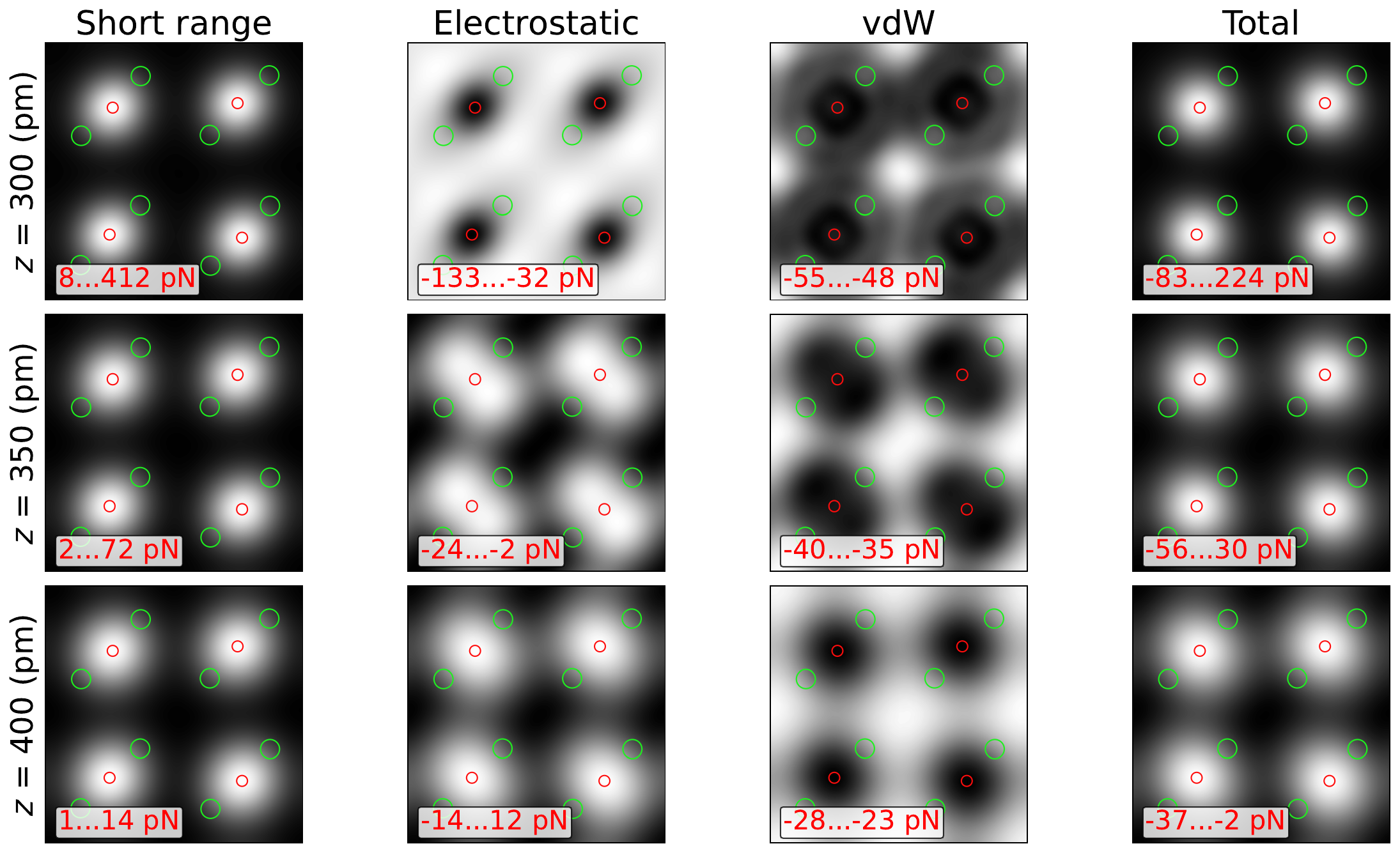}
\includegraphics[width=0.99\textwidth]{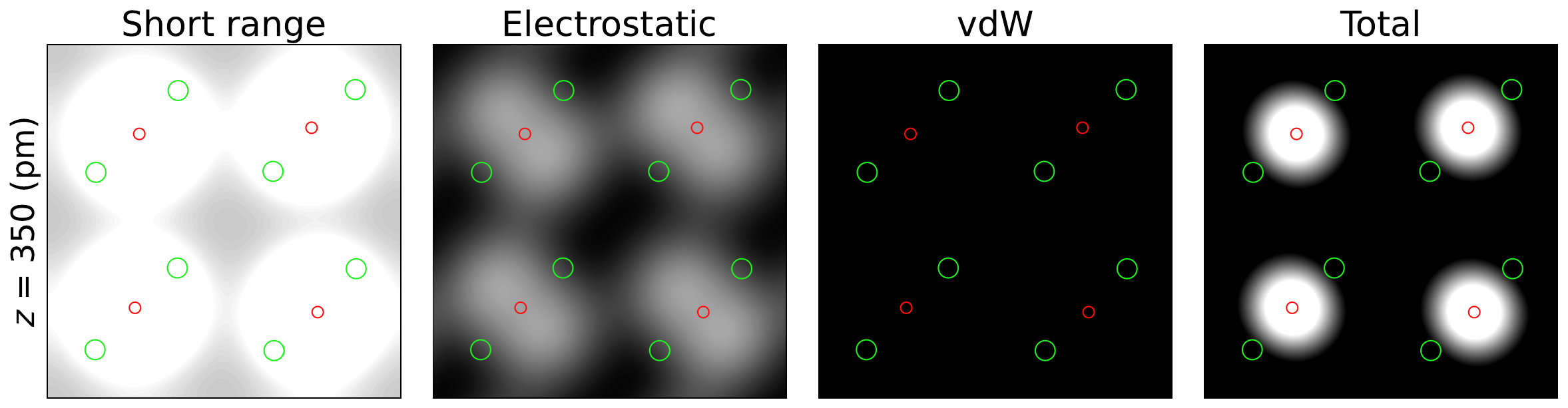}
\caption{
\textbf{FDBM force decomposition for the O-t(100) B2 reconstruction with Ce$^{3+}$ ions confined to the subsurface Ce layer (O-t(100)-top).}
The total force and its short-range (SR), electrostatic (ES), and van der Waals (vdW) components are shown at three probe--surface separations, $z$, referenced to the topmost surface O atoms.
In the upper three rows, each panel is displayed using an independent grayscale, with the minimum and maximum force values indicated as insets.
The bottom row compares the four contributions at $z=350$~pm using a common force scale of $(-25,10)$~pN, where black and white denote the most attractive and least attractive values, respectively.
The contrast is governed primarily by the short-range interaction with the upper O layer.
The electrostatic contribution introduces a weaker modulation that elongates the features along the surface O rows, whereas the vdW term provides an almost spatially uniform attractive background.
}
\label{sfig:SI_fdbm_O_t_subsurf}
\end{center}
\end{figure*}

FDBM decompositions for the CeO$_4$-t(100) reconstruction (Fig.~\ref{sfig:SI_fdbm_2x2}) and the two O-t(100) models \textemdash O-t(100)-mix in Fig.~\ref{sfig:SI_fdbm_O_t_2_layers}, and O-t(100)-top in Fig.~\ref{sfig:SI_fdbm_O_t_subsurf}\textemdash\space reveal markedly different interaction balances between the two surface terminations. At the experimentally relevant imaging heights, the contrast of CeO$_4$-t(100) is largely governed by the electrostatic contribution. By contrast, both O-t(100) models are dominated by the short-range Pauli repulsion associated with the upper O layer, irrespective of whether the Ce$^{3+}$ ions are distributed across the first two Ce layers or confined to the subsurface Ce layer. In these models, the electrostatic interaction provides only a secondary modulation along the surface O rows, while the vdW interaction acts mainly as a nearly uniform attractive background. This difference in the dominant interaction explains why the same simplified probe model performs differently for the two terminations in the force-spectroscopy calculations discussed below.

\clearpage
\subsubsection{Adsorbate models to categorized the X$_p$ feature}

Although the relative weights of the interactions vary among the three adsorbate models considered (H$_\mathrm{2}$O, CO and [CO$_3$]$^{2-}$), the dominant contribution to the AFM contrast always arises from the oxygen atom of the adsorbate, which protrudes with respect to the surface oxygen atoms. This characteristic makes the AFM contrast less sensitive to the extended electrostatic interaction that affects the sub-surface atomic sites of CeO$_\mathrm{4}$-t(100) surface, and therefore a simplified CO-probe model suffices for a qualitative description of the AFM signal.

\begin{figure*}[h!]
\begin{center}
\includegraphics[width=0.99\textwidth]{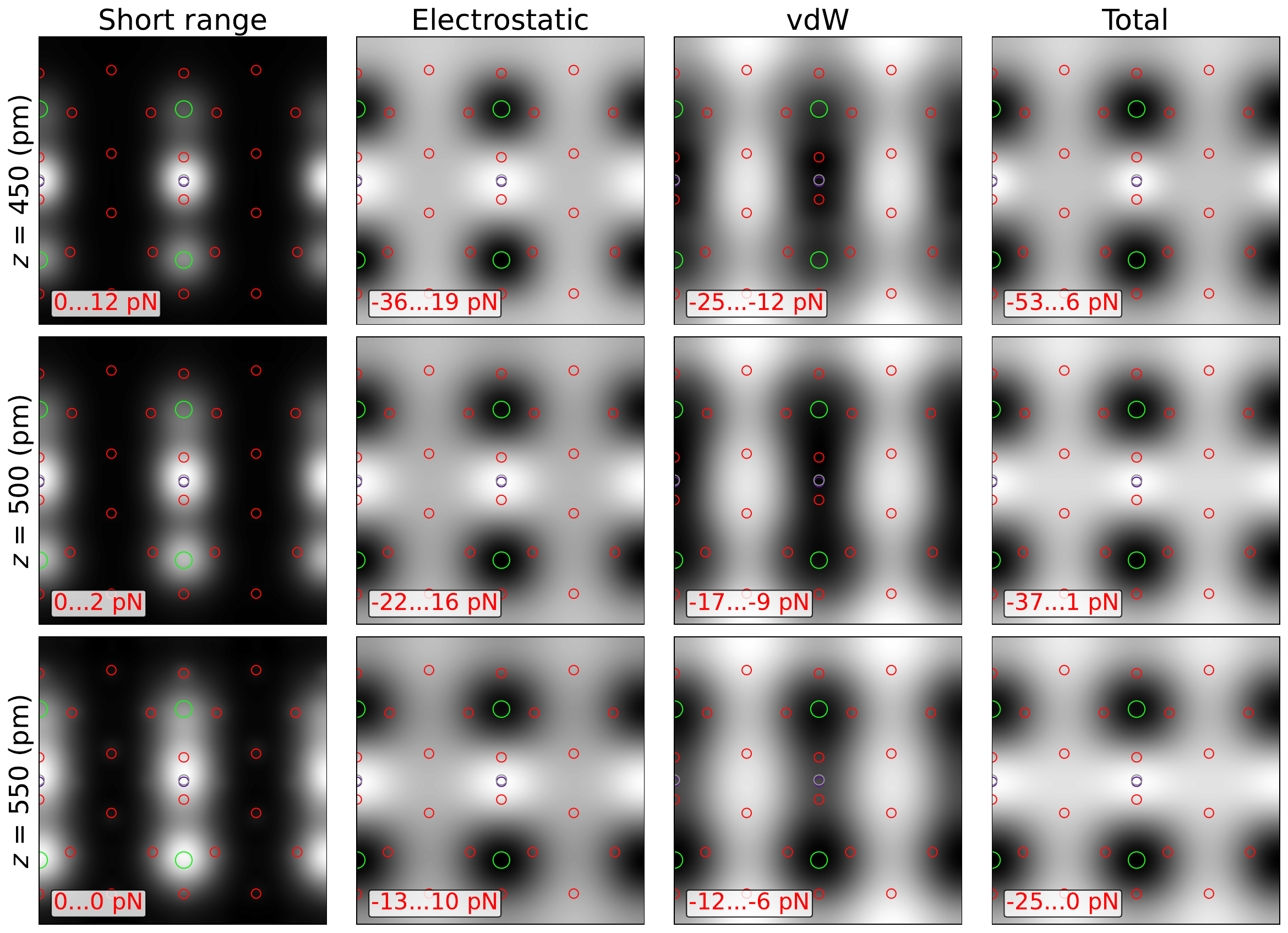}
\includegraphics[width=0.99\textwidth]{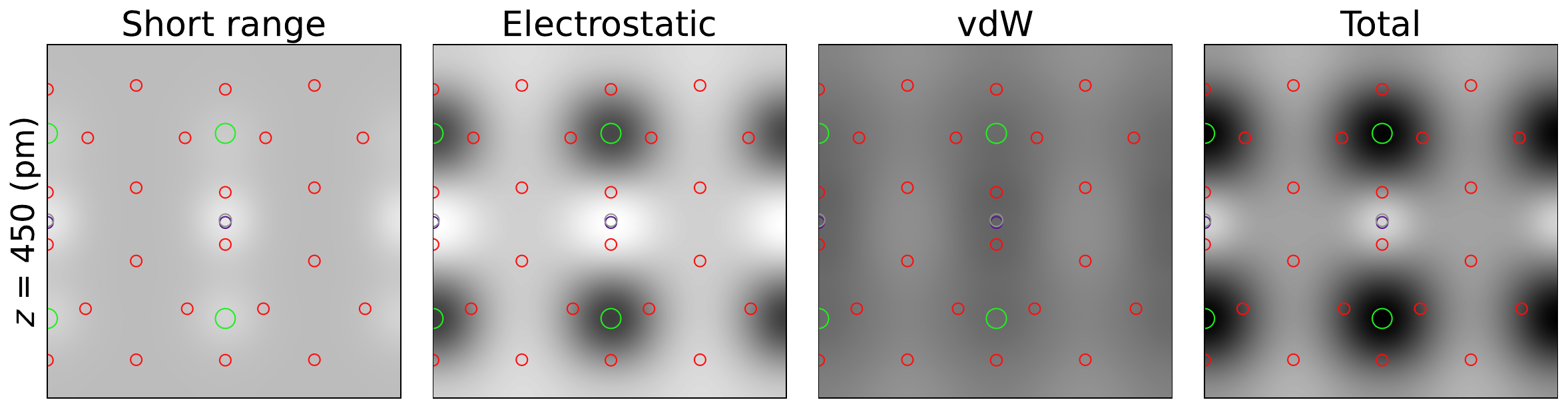}
\caption{
{\bf FDBM force decomposition for a [CO$_3$]$^{2-}$ molecule adsorbed on the CeO$_\mathrm{4}$-t(100) reconstruction.}
Decomposition of the AFM contrast obtained from Full Density-Based Model (FDBM) calculations for a [CO$_3$]$^{2-}$ complex produced by the adsorption of a CO molecule at three probe-surface separations ($z$) measured with respect to the topmost Ce atoms of the pyramids. 
In the configuration considered, the carbon atom interacts with two oxygen atoms of the surface, and the oxygen atom of the CO molecule points outwards (structure in Fig.~\ref{Fig7}A) 
The panels at the bottom row represent the images for $z=450$~pm with the same absolute scale limits for all of them: (black, white) = (-53, 19)~pN.
These values correspond to the minimum (more attractive) and maximum (less attractive) force values found across the four panels.
The contrast on the adsorbate comes from repulsive short-range Pauli and ES contributions compensated by an attractive vdW term.
}
\label{sfig:SI_fdbm_co3}
\end{center}
\end{figure*}

\begin{figure*}[h!]
\begin{center}
\includegraphics[width=0.99\textwidth]{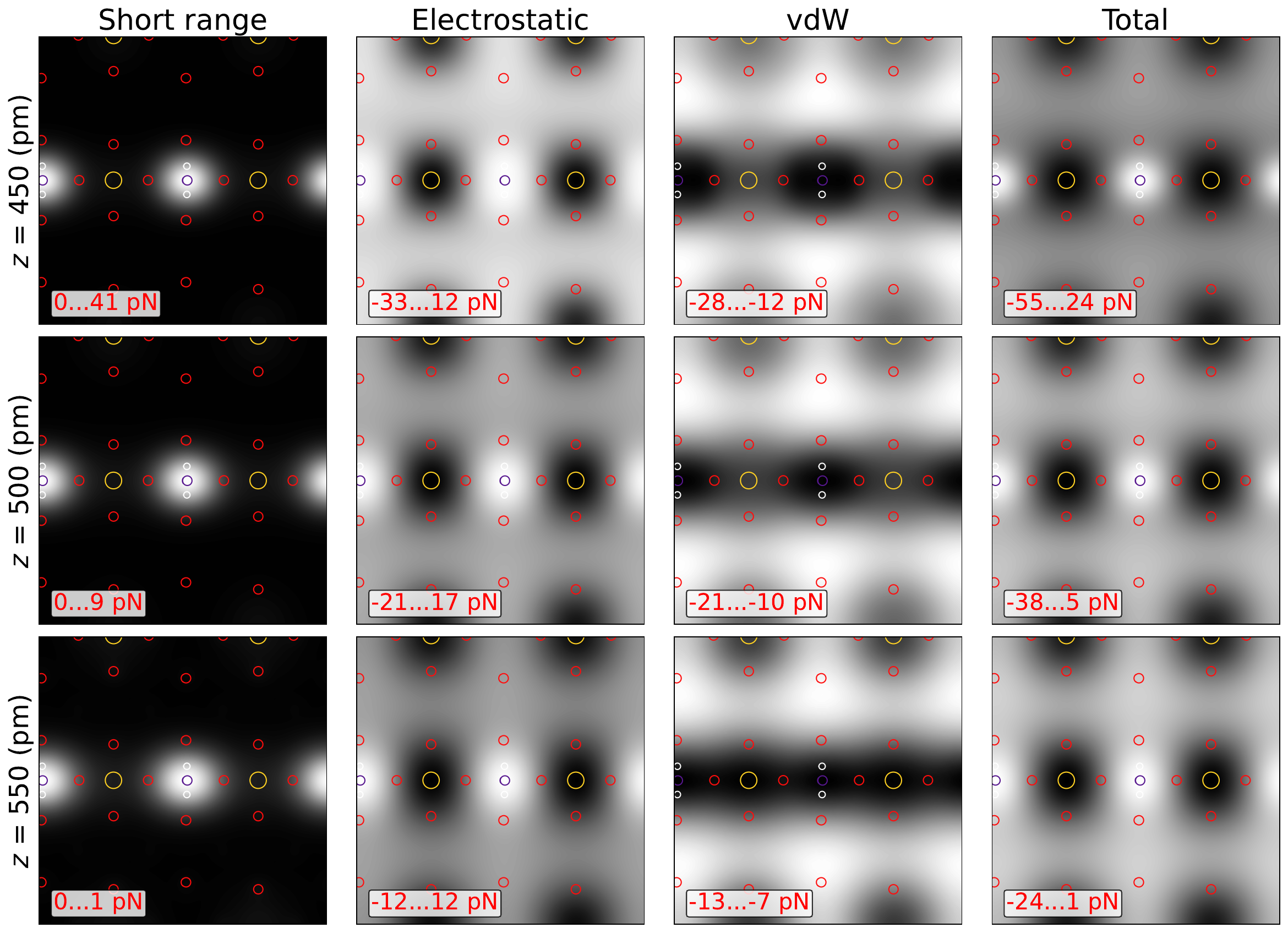}
\includegraphics[width=0.99\textwidth]{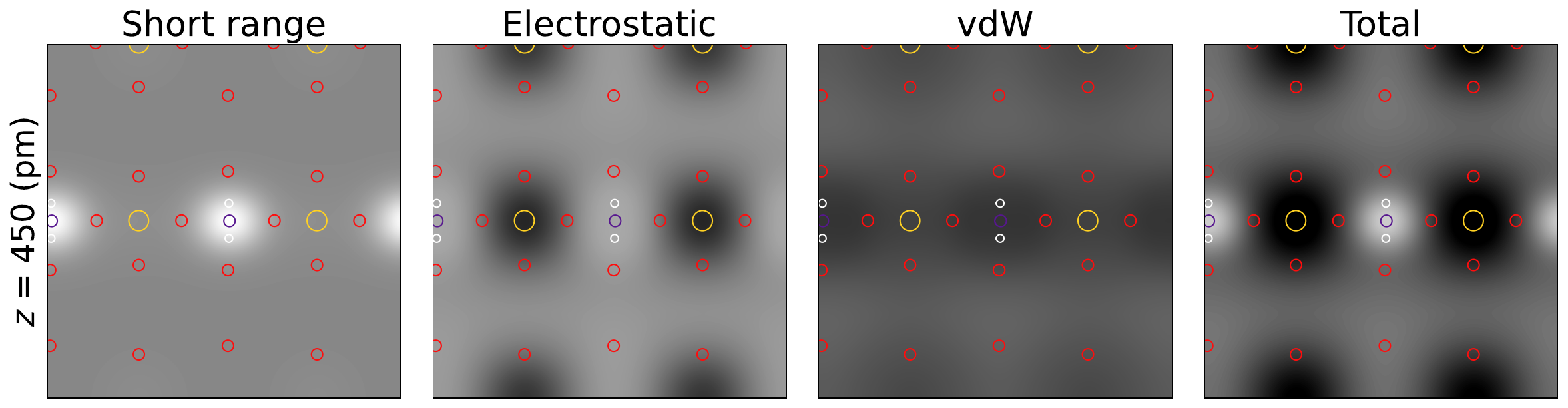}
\caption{
{\bf FDBM force decomposition for a H$_\mathrm{2}$O molecule adsorbed on the CeO$_\mathrm{4}$-t (100) surface.}
Decomposition of the AFM contrast obtained from Full Density-Based Model (FDBM) calculations for the H$_\mathrm{2}$O molecule, forming H-bonds with two oxygen atoms of the surface. 
In this configuration, the H$_\mathrm{2}$O's oxygen is pointing outwards (structure in Fig.~\ref{Fig7}B). 
The top three rows correspond to three probe-surface separations ($z$) measured with respect to the position of the topmost Ce atoms of the pyramids.
The total AFM force (Total) is composed of short-range Pauli repulsion (SR), electrostatic (ES), and van der Waals (vdW) contributions. 
For the top three rows, each of the panels uses its own gray scale, with the minimum (black) and maximum (white) force values displayed as insets.
The panels at the bottom row represent the images for $z=450$~pm with the same absolute scale limits for all of them: (black, white) = (-55, 41)~pN.
These values correspond to the minimum (more attractive) and maximum (less attractive) force values found across the four panels.
The contrast on the adsorbate comes from repulsive short-range Pauli and ES contributions compensated by an attractive vdW term.
}
\label{sfig:SI_fdbm_h2o}
\end{center}
\end{figure*}

\begin{figure*}[t!]
\begin{center}
\includegraphics[width=0.99\textwidth]{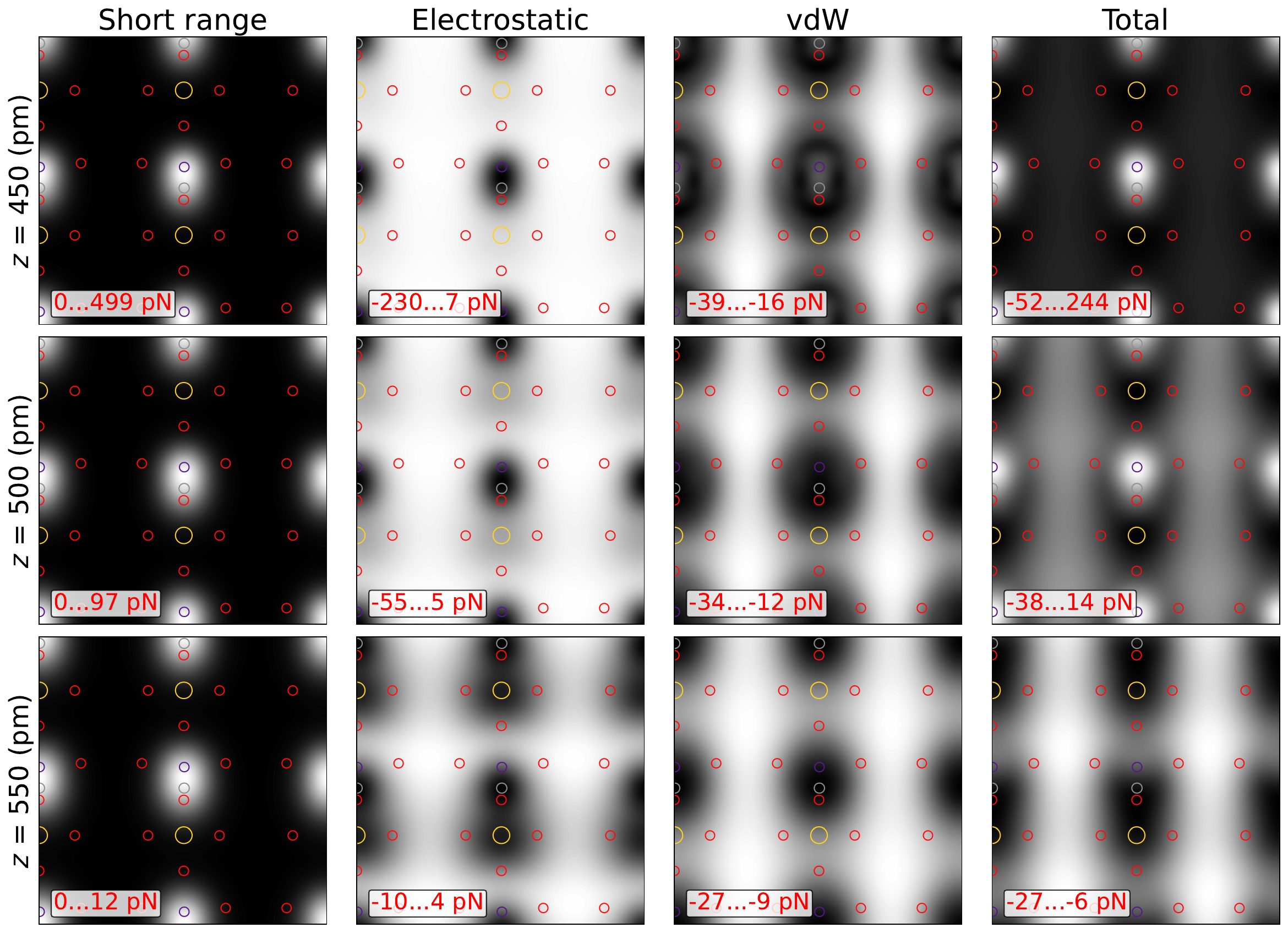}
\includegraphics[width=0.99\textwidth]{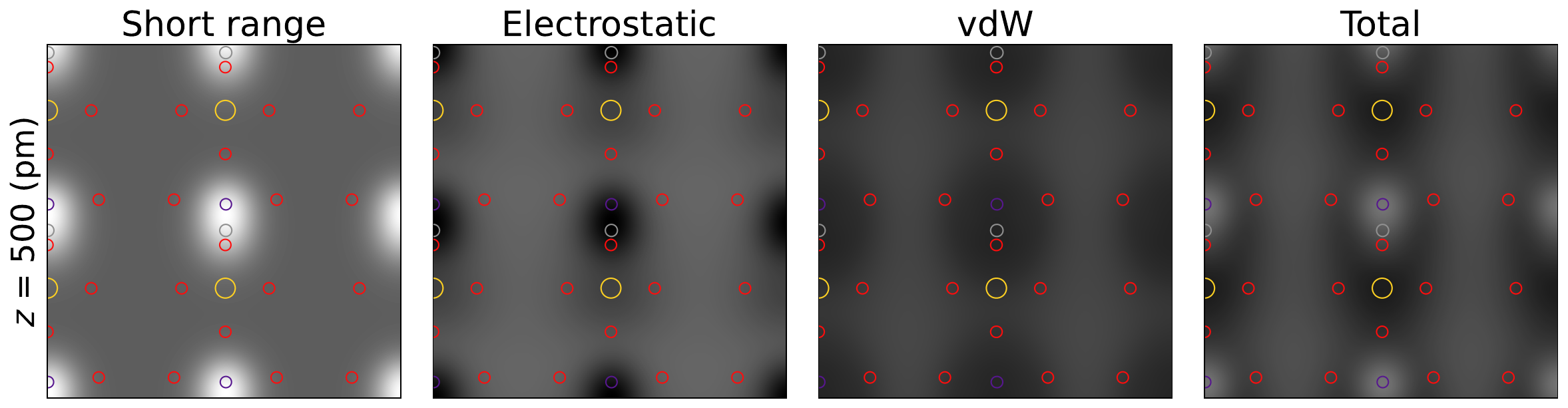}
\caption{
{\bf FDBM force decomposition for a CO molecule adsorbed on the CeO$_\mathrm{4}$-t(100) surface.}
Decomposition of the AFM contrast obtained from Full Density-Based Model (FDBM) calculations for a CO molecule adsorbed on the topmost Ce atom of a CeO$_\mathrm{4}$-pyramid.
In this configuration, the CO is tilted towards a Ce-Ce bridge position (structure in Fig.~\ref{Fig7}C).
The top three rows correspond to three probe-surface separations ($z$) measured with respect to the position of the topmost Ce atoms of the pyramids.
The panels at the bottom row represent the images for $z=500$~pm with the same absolute scale limits for all of them: (black, white) = (-55, 97)~pN.
These values correspond to the minimum (more attractive) and maximum (less attractive) force values found across the four panels.
The contrast on the adsorbate comes from a combination of repulsive SR and attractive ES and vdW contributions.
}
\label{sfig:SI_fdbm_co}
\end{center}
\end{figure*}

\clearpage

\subsection{Deviation of the calculated $\Delta$f(z)  from the experiments: the quest for a CuO$_X$ probe model}

For the CeO$_\mathrm{4}$-t(100) surface, the calculated $\Delta$f(z) curve over the Ce atom atop the pyramid provides a good quantitative description of both the shape and the value of the minimum, whereas the calculated curves over the coordination vacancy and a position between two oxygen atoms of adjacent CeO$_\mathrm{4}$ pyramids deviate from the experiments. The site dependence of this discrepancy indicates that the problem is linked not only to the probe model but also to the dominant interaction at each surface position.

A simple probe model such as the one used in this work (a rigid CO molecule with its oxygen atom pointing towards the surface), reproduces reasonably well features dominated by short-range Pauli repulsion, whose exponential decay makes the interaction mainly controlled by the lone pair of the probe. This is the situation in the case of the CeO$_2$(111) surface, where the same type of model to simulate the probe has provided quantitative agreement with the experiments~\cite{oscar_manuel_ceria_111_water}. 
For the CeO$_\mathrm{4}$-t(100) surface, however, the FDBM decomposition shows that, at probe--surface separations relevant to the experiment, the AFM contrast is largely governed by the electrostatic contribution (Fig.~\ref{sfig:SI_fdbm_2x2}) which is longer-ranged and therefore more sensitive to the detailed charge distribution of the whole apex.
The curve over the Ce atoms at the pyramids \textemdash which protrude about 100~pm above the surrounding oxygen atoms (Fig.~\ref{sfig:SI_DFT_Niklas})\textemdash\space is reasonably well described by a simple CO probe model, because the interaction is localized at the foremost atom of the probe and the contribution from neighboring atoms of the apex becomes less relevant.

\begin{figure}[t!]
    \centering
    \includegraphics[width=0.8\linewidth]{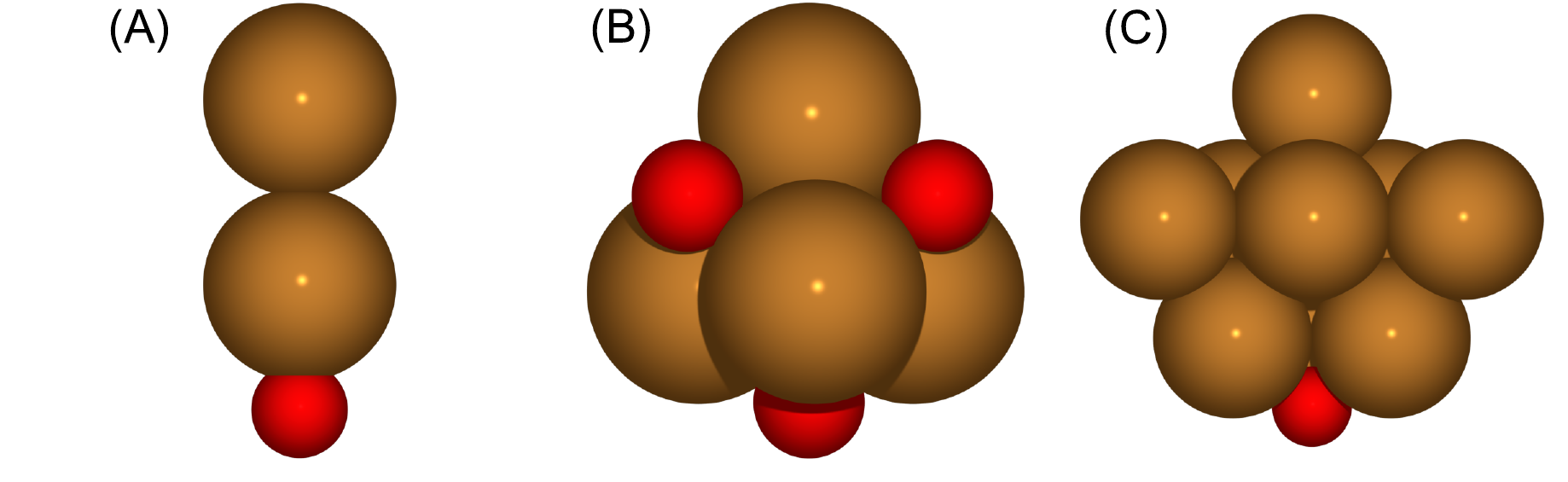}
    \caption{
\textbf{
Geometries of Cu$_x$O clusters tested as alternative probe models to a rigid CO molecule}.
(A) Cu$_2$O, (B) Cu$_4$O$_4$, and (C) Cu$_{10}$O. Structures correspond to DFT-relaxed clusters and are shown in the orientation used for the force-spectroscopy simulations. Copper and oxygen atoms are shown in brown and red, respectively.
}
\label{sfig:copper_oxide_tips}
\end{figure}

To try to improve the agreement between experimental and calculated force curves over the coordination vacancy and over a position mid-point between oxygen atoms of adjacent pyramids, we also considered several CuO$_x$ probe models (Fig.~\ref{sfig:copper_oxide_tips}). However, all of them yielded maximum attractive forces significantly larger than those obtained with a rigid CO probe, in some cases by up to a factor of six. Thus, increasing the structural complexity of the apex did not lead to a quantitative improvement in the description of the experimental curves.

In contrast, for the O-t(100) surface, the FDBM decomposition shows that the AFM signal is dominated by short-range Pauli repulsion, with only a modest electrostatic modulation along the oxygen rows (Fig.~\ref{sfig:SI_fdbm_O_t_subsurf}). In this case, the simplified probe model reproduces both the AFM image contrast and the force-spectroscopy curves satisfactorily.

For the adsorbate models considered to categorize the X$_p$ feature, the FDBM decomposition indicates different balances between interaction channels: the carbonate case is largely electrostatic; water contains important short-range and electrostatic contributions; and, the CO at a Ce site is mainly governed by short-range repulsion (Figs.~\ref{sfig:SI_fdbm_co3}, \ref{sfig:SI_fdbm_h2o}, and~\ref{sfig:SI_fdbm_co}). Despite these differences, in all three cases the protruding oxygen atom of the adsorbate is the foremost interacting species, so the simplified probe is expected to provide a more robust qualitative description than for the sub-surface sites of CeO$_\mathrm{4}$-t(100) surface.

Overall, our results indicate that simple probe-particle or rigid-apex~\cite{Lammers2021, Monig2018-probeparticle}  descriptions can be sufficient for a qualitative interpretation of CuO$_x$-functionalized AFM contrast on sites dominated by localized short-range interactions, but may become inadequate for polar and corrugated oxide surfaces where long-range electrostatics and the detailed charge distribution of the apex play a central role. A realistic quantitative model of copper-oxide-functionalized probes therefore remains an open problem.

\section{STM simulations, projected density of states (pDOS) and isosurfaces for the interpretation of the STM contrast} \label{secSI:stm_simulations}

Within the Tersoff--Hamann approximation~\cite{TersoffHamann}, the simulated STM signal is proportional to the integrated local density of states in the vacuum region, so the contrast reflects the spatial decay of the electronic states into vacuum rather than the atom-projected density of states (DOS) alone.

The isosurfaces shown in this section correspond to constant values of the integrated local DOS and provide a qualitative representation of the tunneling current. Unless otherwise stated, the isovalues differ from those used in the main text, and they were adjusted to enhance the visibility of the relevant features.
As a result, maxima in the STM contrast do not necessarily coincide with the atomic sites carrying the largest projected electronic weight, but may instead reflect contributions from neighboring atoms or deeper layers, as discussed below.

\begin{figure}[b!]
    \centering
    \includegraphics[width=0.9\linewidth]{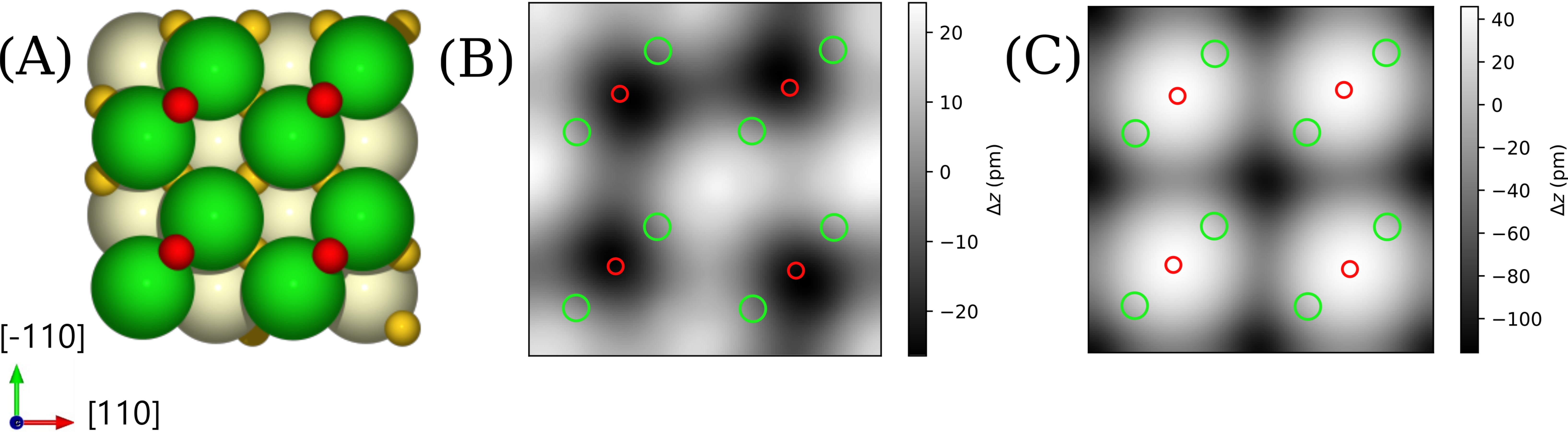}
\caption{%
\textbf{Simulated constant-current STM images of the reduced O-t(100) surface with Ce$^{3+}$ ions confined to the subsurface Ce layer (O-t(100)-top).}
(\textbf{A}) Structural model of the O-t(100)-top surface.
(\textbf{B}, \textbf{C}) Simulated empty- and filled-state STM images, respectively.
The filled-state maxima are centered on the top-layer oxygen atoms, consistent with the dominant contribution of their occupied states.
In the empty-state image, the maxima are displaced from the uppermost oxygen sites and are instead modulated by the vacuum decay of unoccupied states associated predominantly with the second oxygen layer, as shown by the pDOS and isosurfaces in Fig.~\ref{sfig:SI_stm_pdos_ot_subsurf}.
Isovalues correspond to the mean value in a plane 3~$\mathring{\mathrm{A}}$ above the top oxygen layer. The empty- and filled-state isovalues are $1.49\times10^{-5}$ and $6.31\times10^{-5}$~states~$\mathring{\mathrm{A}}^{-3}$, respectively.
}
\label{sfig:SI_stm_ot_subsurf}
\end{figure}

\begin{figure}
    \centering
    \includegraphics[width=0.99\linewidth]{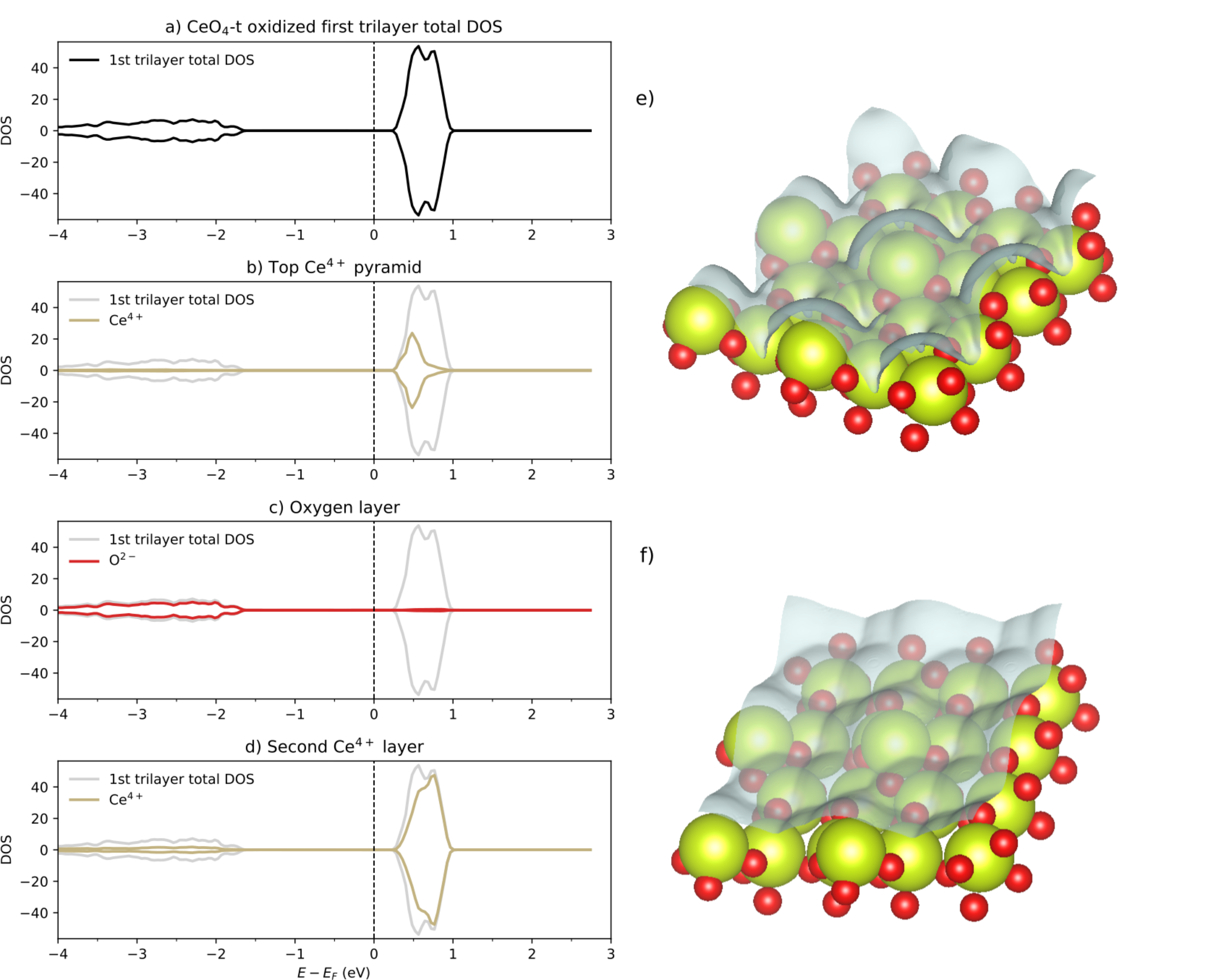}
    \caption{
    \textbf{pDOS and STM isosurfaces for the oxidized CeO$_\mathrm{4}$-t(100) model.}
    (A) Total pDOS of the first trilayer. 
    (B--D) Contributions from the top Ce$^{4+}$ atom of the CeO$_4$ pyramid, the surface oxygen layer, and the second Ce$^{4+}$ layer, respectively. 
    (E,F) Isosurfaces of the integrated local DOS used for the empty- and filled-states STM simulations, respectively.
    In the empty-states window, the tunneling density is governed by Ce$^{4+}$-derived unoccupied states.
    The weak annular character around the top Ce$^{4+}$ site reflects a lateral redistribution of the vacuum local DOS by the surrounding oxygen environment rather than a simple atom-centered spherical state. %
    In filled-states, the pDOS is dominated by the oxygen layer below the top cation, while the top Ce$^{4+}$ atom carries negligible projected weight. Nevertheless, the filled-states isosurface (F) places the apparent maximum above the top Ce$^{4+}$ site, showing that the tunneling contrast is governed by the vacuum tail of the integrated local DOS rather than by the atom-projected DOS alone.
    The isovalues used to render (E,F) are ten times the corresponding values used in 
    Figs.~\ref{Fig3}B and~\ref{Fig3}C, facilitating visualization of the three-dimensional features.
    }
    \label{sfig:SI_stm_pdos_niklas_ox}
\end{figure}

\begin{figure}
    \centering
    \includegraphics[width=0.99\linewidth]{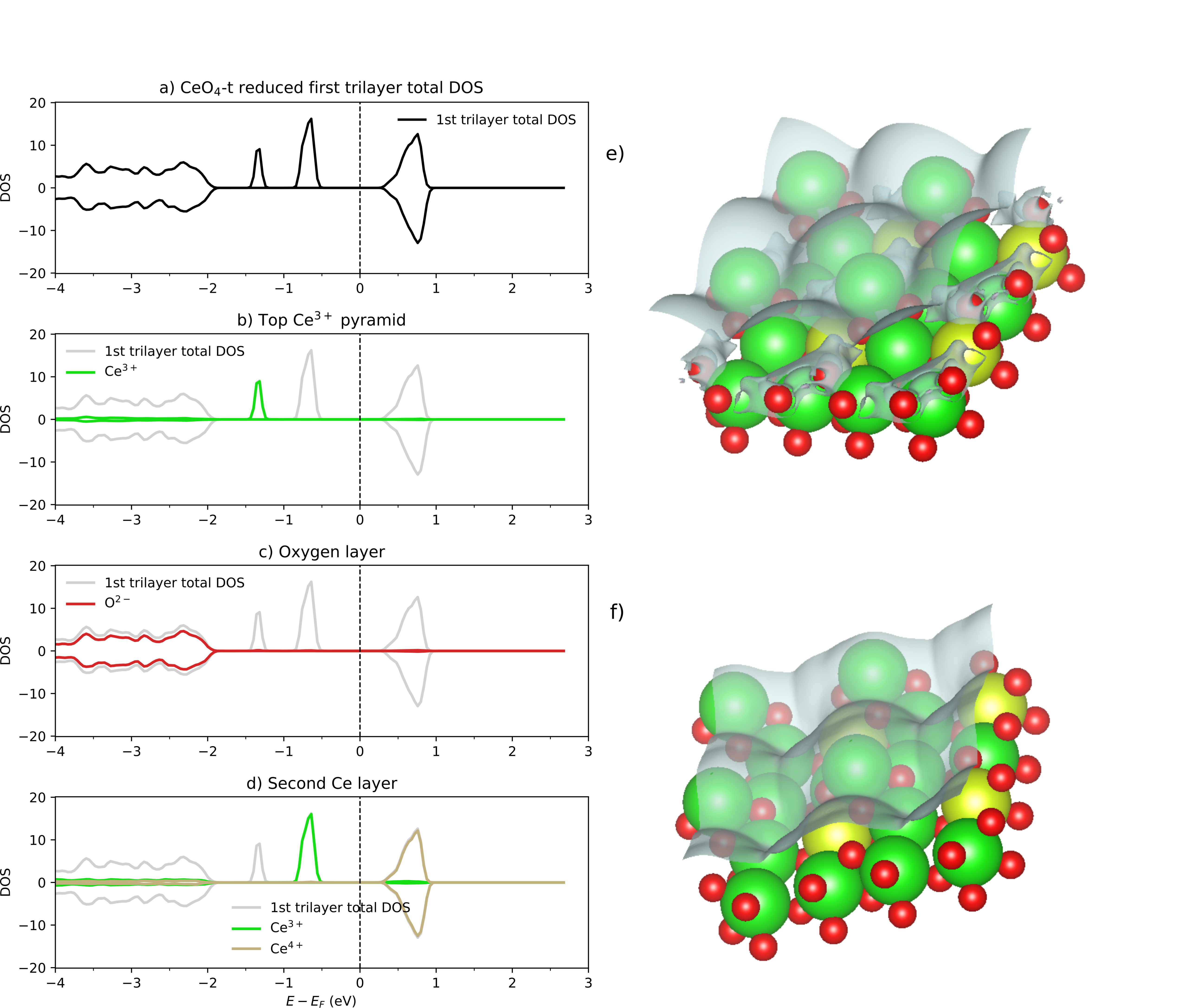}
    \caption{
    \textbf{pDOS and STM isosurfaces for the reduced CeO$_\mathrm{4}$-t(100) model.}
    (A) Total pDOS of the first trilayer. 
    (B--D) Contributions from the top Ce$^{3+}$ atom of the CeO$_4$ pyramid, the surface oxygen layer, and the second Ce layer, respectively. 
    (E,F) Isosurfaces of the integrated local DOS used for the empty- and filled-states STM simulations, respectively. 
    The occupied states contain broad O-derived contributions together with a localized Ce$^{3+}$ peak at the top pyramid atom, consistent with a filled-states protrusion that becomes asymmetric because of the nonequivalent Ce$^{3+}$/Ce$^{4+}$ environment. 
    In empty-states , the pDOS is dominated by Ce$^{4+}$-derived states, while the Ce$^{3+}$ ions contribute negligibly in this energy range.
    The pronounced cap visible above the top Ce$^{3+}$ position in empty-states is therefore not supported by the pDOS as a clearly localized empty state on that atom. A more plausible interpretation is that this feature corresponds to a vacuum maximum arising from neighboring Ce$^{4+}$-derived unoccupied states and the interstitial redistribution of the local DOS. Because the STM signal is determined from the vacuum local DOS whereas the pDOS is evaluated within finite atom-centered projection spheres, a minor contribution from a highly extended state above the top Ce$^{3+}$ site cannot be completely excluded. %
    The isovalues used to render (E,F) are ten times the corresponding values used in Figs.~\ref{Fig3}E and ~\ref{Fig3}F, facilitating visualization of the three-dimensional features.}
    \label{sfig:SI_stm_pdos_niklas_red}
\end{figure}

\begin{figure}
    \centering
    \includegraphics[width=0.9\linewidth]{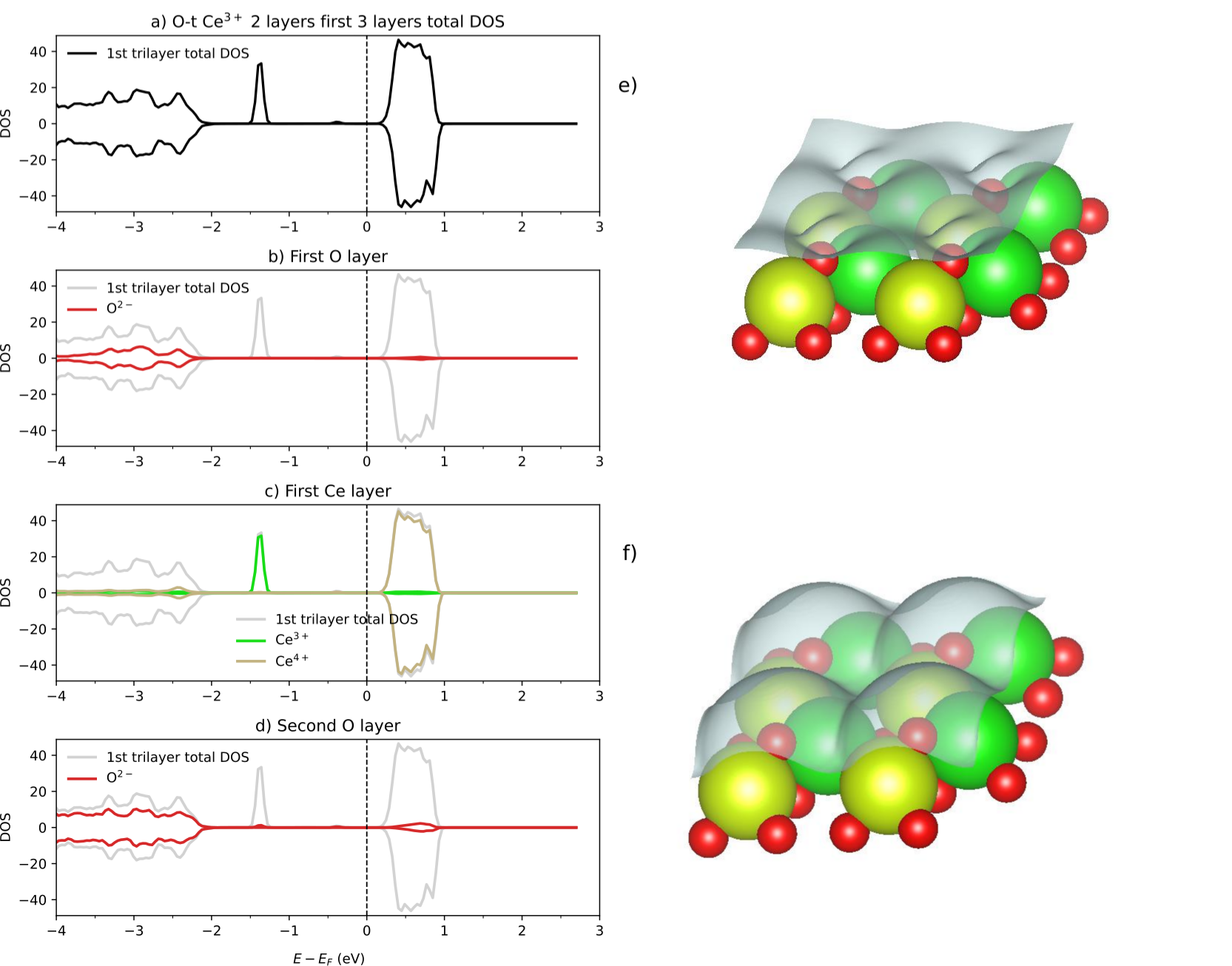}
    \caption{
    \textbf{pDOS and STM isosurfaces for the O-t(100) model with Ce$^{3+}$ atoms distributed within the first two Ce layers (O-t(100)-mix).}
    (A) Total pDOS of the first three layers. 
    (B--D) Contributions from the first O layer, the first Ce layer, and the second O layer, respectively. 
    (E,F) Isosurfaces of the integrated local DOS used for the empty- and filled-states STM simulations, respectively. 
    As in the O-t(100)-top model (Fig.~\ref{sfig:SI_stm_pdos_ot_subsurf}), the occupied signal is mainly O-derived and the filled-states STM contrast is therefore dominated by the top oxygen layer.
    By contrast, the empty-states pDOS is overwhelmingly governed by the top-layer Ce$^{4+}$ site, whereas the Ce$^{3+}$ and O contributions are much smaller. 
    The corresponding empty-states isosurface develops a nonspherical, off-centered lobe above the top Ce$^{4+}$ atom, displaced away from the nearest upper O site. 
    This indicates that the unusual asymmetric empty-states STM contrast originates from the shape and vacuum decay of the Ce$^{4+}$-derived unoccupied density rather than from a simple topographic corrugation.
        The isovalues used to render (E,F) are ten times the corresponding values used in Figs.~\ref{Fig6}B and~\ref{Fig6}C, facilitating visualization of the three-dimensional features.}
    \label{sfig:SI_stm_pdos_ot_2layers}
\end{figure}

\begin{figure}
    \centering
    \includegraphics[width=0.99\linewidth]{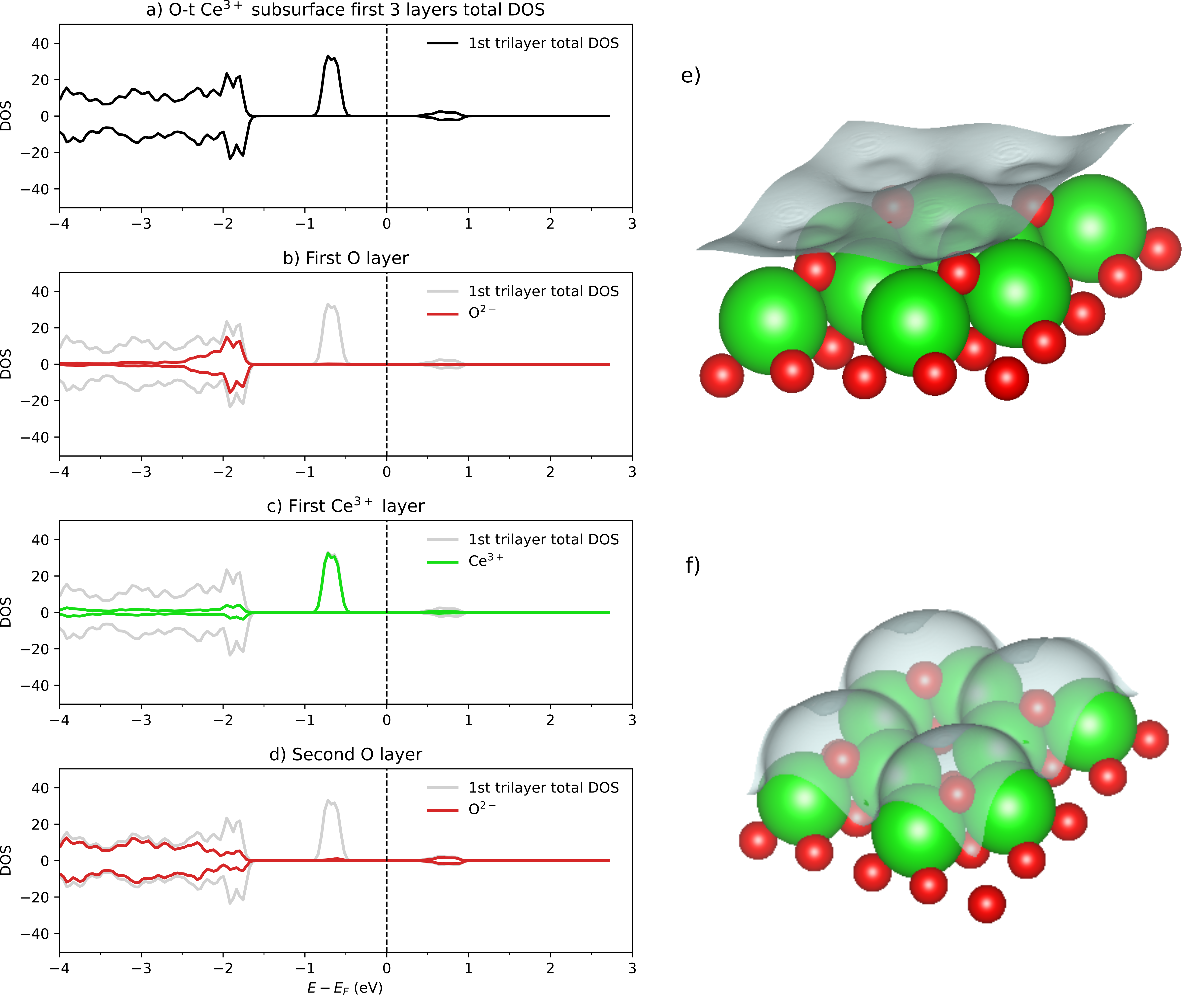}
    \caption{
    \textbf{pDOS and STM isosurfaces for the O-t(100) model with subsurface Ce$^{3+}$ (O-t(100)-top).}
    (A) Total pDOS of the first three layers. 
    (B--D) Contributions from the first O layer, the first Ce$^{3+}$ layer, and the second O layer, respectively. 
    (E,F) Isosurfaces of the integrated local DOS used for the empty- and filled-states STM simulations, respectively. 
    The occupied signal is dominated by O-derived states from the upper oxygen layer, consistent with maxima centered on the top O atoms in the filled-states STM image. 
    In the empty-states window, the first O layer and the subsurface Ce$^{3+}$ site carry only a minor contribution, whereas the second O layer provides the main O-derived unoccupied weight.
     As a consequence, the maxima of the empty-states isosurface are not centered on the uppermost O atoms, but are modulated by the second oxygen layer, showing that the STM contrast is controlled by the vacuum decay of deeper O-derived states rather than simply by the highest atomic positions.
    The isovalues used to render (E,F) are ten times the corresponding values used in Figs.~\ref{sfig:SI_stm_ot_subsurf}B and~\ref{sfig:SI_stm_ot_subsurf}C, facilitating visualization of the three-dimensional features. }
    \label{sfig:SI_stm_pdos_ot_subsurf}    
\end{figure}

\clearpage

\bibliography{Updated_bib,additional}